IAPS | Institute for AI Policy and Strategy

June 2024

# Coordinated Disclosure of Dual-Use Capabilities: An Early Warning System for Advanced AI

By Joe O'Brien, Shaun Ee, Jam Kraprayoon, Bill Anderson-Samways, Oscar Delaney, and Zoe Williams

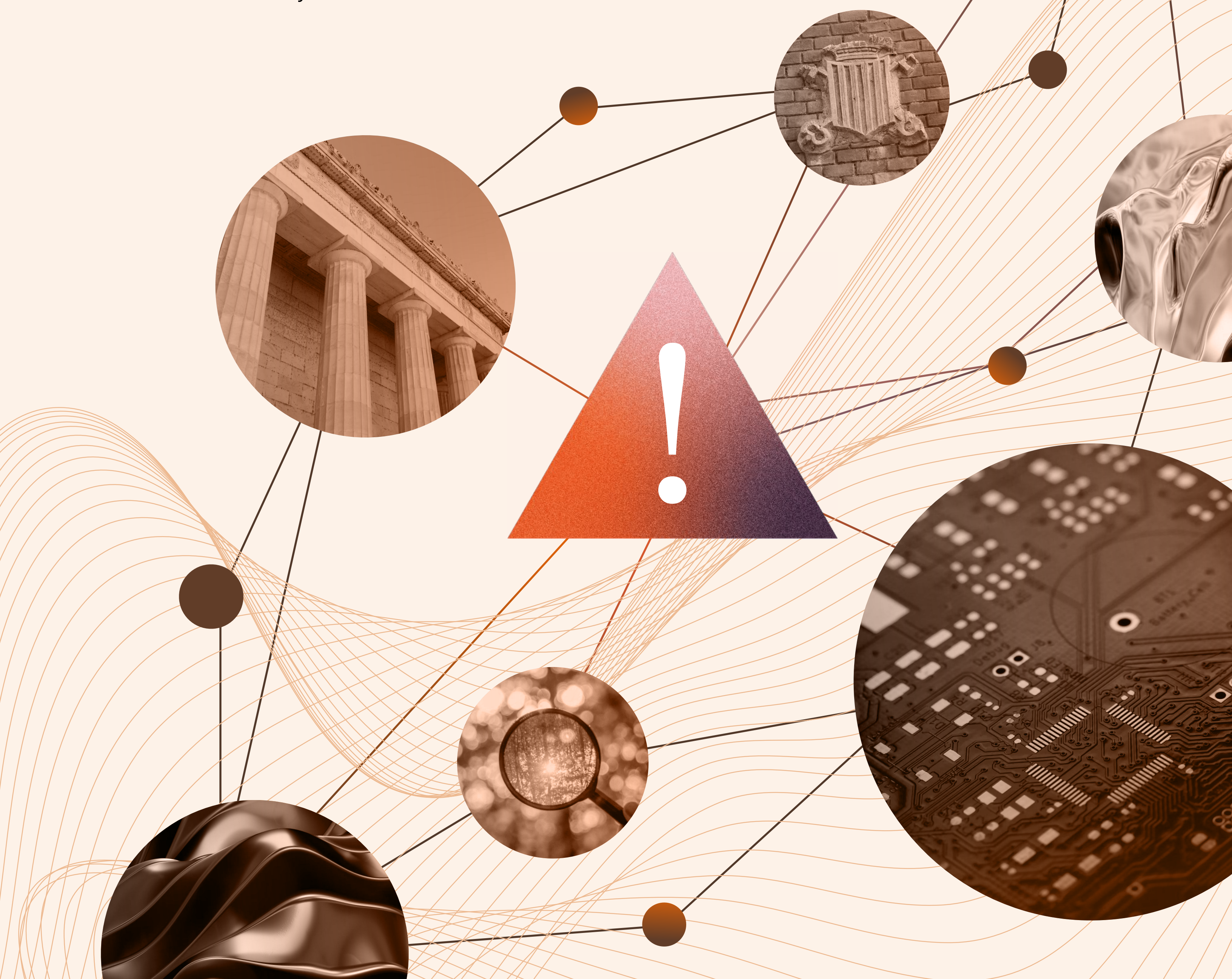

# Abstract

Advanced AI systems may be developed which exhibit capabilities that present significant risks to public safety or security. They may also exhibit capabilities that may be applied defensively in a wide set of domains, including (but not limited to) developing societal resilience against AI threats.

We propose **Coordinated Disclosure of Dual-Use Capabilities** (CDDC) as a process to guide early information-sharing between advanced AI developers, US government agencies, and other private sector actors about these capabilities. The process centers around an information clearinghouse (the "coordinator") which receives evidence of dual-use capabilities from finders via mandatory and/or voluntary reporting pathways, and passes noteworthy reports to defenders for follow-up (i.e., further analysis and response). This aims to provide the US government, dual-use foundation model (DUFM) developers, and other actors with an overview of AI capabilities that could significantly impact public safety and security, as well as maximal time to respond.

We make several recommendations:

1. Congress should assign a coordinator within the US government to receive and distribute reports on dual-use AI capabilities ("DUC reports"). This should be paired with strengthened reporting requirements for DUCs.
2. Either the President via Executive Order or Congress via legislation should assign agency leads for working groups of "defender" agencies—agencies that receive DUC reports from the coordinator and act on them.
3. Congress should fund the US AI Safety Institute to build capacity for wider government involvement in model evaluations (by enabling agencies to directly perform evaluations, or audit or otherwise be involved in company-run evaluations).
4. National Institute of Standards and Technology (NIST) or alternatively, a non-governmental organization such as Carnegie Mellon University's Software Engineering Institute (CMU SEI) or the Frontier Model Forum (FMF) should lead efforts with AI developers, relevant agencies, and third parties to develop common language for DUC reporting and triage.
5. DUFM developers should establish clear policies and intake procedures for independent researchers reporting dual-use capabilities, based on Vulnerability Reporting Policies.
6. DUFM developers should maintain incident response plans for DUCs and build working relationships with defenders in government and relevant non-governmental organizations.
7. DUFM developers should collaborate with working groups (once such groups are developed) to identify capabilities that could help defenders, which can be shared via the CDDC infrastructure.

# Executive Summary

**Society needs a secure, clear reporting system for information regarding dual-use AI capabilities** (DUCs) to give governments, AI companies, and other defense-relevant actors time to mitigate AI risks. AI experts including prominent scholars such as Geoffrey Hinton and Yoshua Bengio (Bengio et al., 2024) and CEOs from major AI companies Google DeepMind, OpenAI, and Anthropic (Roose, 2023b) have affirmed the need for society to manage risks from advanced AI systems, in line with similar affirmations from a number of governments (Department for Science, Innovation & Technology, 2024e). These risks span an extremely broad range of scenarios, and range from already evidenced to speculative. They include the use of AI to enable cybercrime (Hamin & Scott, 2024) and biological attacks (Mouton et al., 2024), weaken our shared understanding of reality (Bengio et al., 2024), and also include loss-of-control scenarios (Bengio, 2024, pp. 51–53), among other concerns.

**However, there are several notable gaps in the existing AI governance infrastructure that limit the use and efficacy of capabilities reporting, including:**

1. An *evaluation gap*: there is a lack of a robust evaluation system, including both (a) technical evaluations, and (b) agreement on what capabilities pose dangers to public safety and/or security.[1]

2. A *reporting gap*: several disincentives to reporting information to government encourage non-reporting, across multiple types of finders.

3. A *coordination gap*: we find challenges with the existing mandatory capabilities reporting system in the US established under Executive Order 14110 Sec. 4.2(a)(i) that limit the US federal government's potential use of capabilities information.

4. A *defense gap:* there is a lack of clear ownership for specific risk areas, especially for AI-specific emerging risks (such as model autonomy or deception).

**To address some of these gaps with the aim of establishing more effective management of advanced AI capabilities and their risks, this document proposes Coordinated Disclosure of Dual-Use Capabilities (CDDC).** CDDC is a process to guide information-sharing about DUCs between model developers, US government agencies, and other private sector actors, based loosely on the concept of coordinated vulnerability disclosure (CVD) in cybersecurity.

[1] In this paper we do not attempt to solve this evaluation gap; we take as given that evaluation techniques exist and will be improved upon.

**CDDC describes a set of reporting pathways from finders, through a coordinator, to defenders,** optimized to allow defenders maximal time to respond to the existence of a given capability before information about the capability is made public (Map 1). The process centers around an information clearinghouse (the "*coordinator*") which receives evidence of DUCs from *finders* via mandatory and/or voluntary reporting pathways, and passes noteworthy reports to *defenders* for follow-up (i.e., further analysis and response). This aims to provide the government and dual-use foundation model (DUFM) developers with a comprehensive overview of AI capabilities that could significantly impact public safety and security, and give them time to respond and implement countermeasures by providing early and private access to information about DUCs.

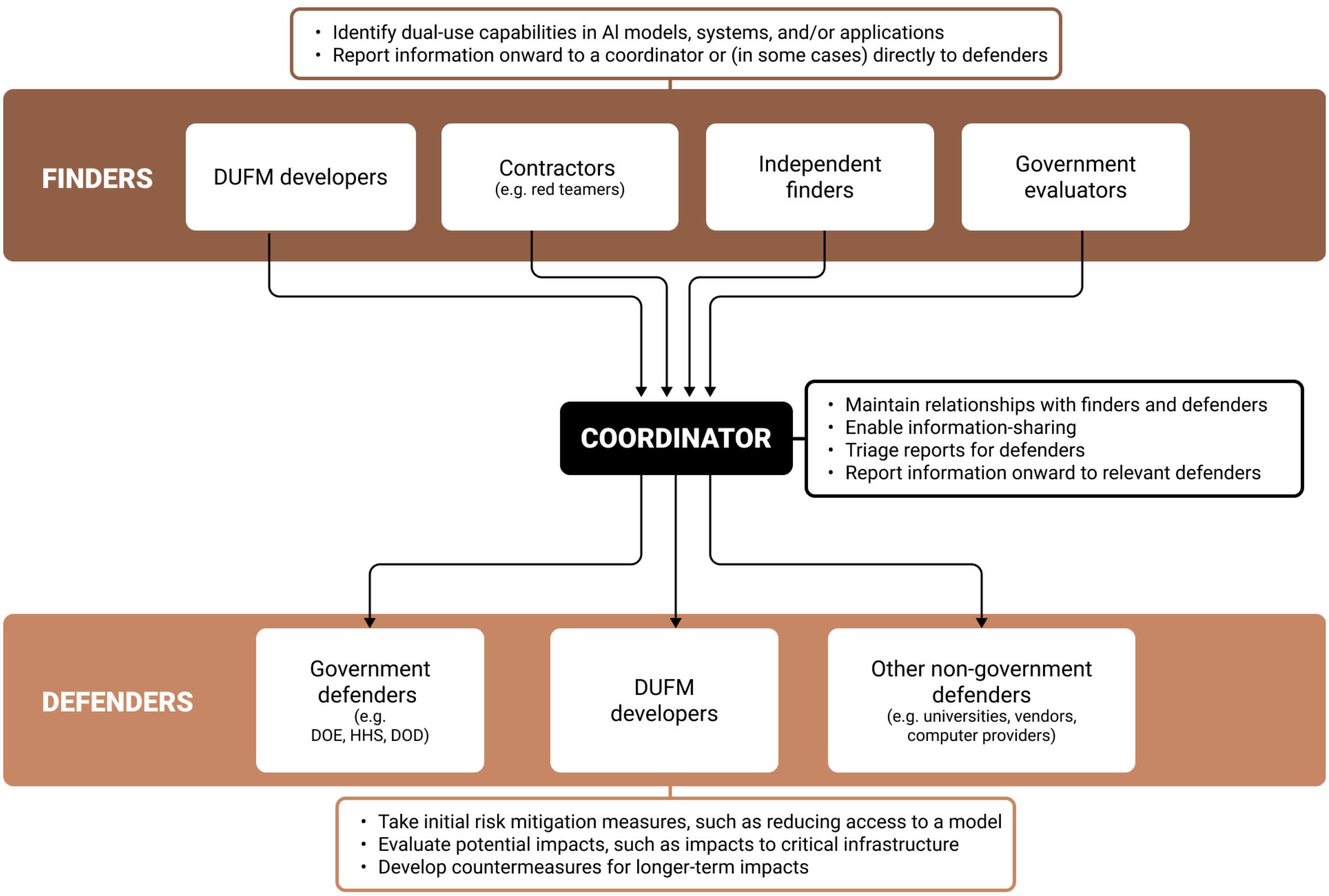

*Map 1. Basic Structure of CDDC* [2]

[2] Importantly, the suggested structure is not designed to preclude other forms of information-sharing; this is primarily intended to provide a central coordination point as a floor (rather than a ceiling) to ensure the

For a definition of "dual use capabilities", we lean on the definition of Dual-Use Foundation Models provided in the Executive Order on Safe, Secure, and Trustworthy Development of Artificial Intelligence (The White House, 2023b), which describes the capabilities of such models as: "[...] high levels of performance at tasks that pose a serious risk to security, national economic security, national public health or safety, or any combination of those matters, such as by: (i) substantially lowering the barrier of entry for non-experts to design, synthesize, acquire, or use chemical, biological, radiological, or nuclear (CBRN) weapons; (ii) enabling powerful offensive cyber operations through automated vulnerability discovery and exploitation against a wide range of potential targets of cyber attacks; or (iii) permitting the evasion of human control or oversight through means of deception or obfuscation." We also argue that a more precise definition for dual-use capabilities could provide benefits across the AI governance ecosystem.

We focus on reporting for dual-use capabilities (DUCs) because they pose an urgent and serious risk to security and public safety, but also because establishing better information-sharing and coordination around DUCs may have additional benefits—such as:

- Providing insight into how capabilities of AI systems more broadly will develop in the future.
- Alerting defense-relevant actors to defensive uses of dual-use capabilities.
- Strengthening visibility of AI capabilities to support better-informed AI governance.

We argue that there are concrete actions that defenders can take when given information on dual-use capabilities, and that reporting is therefore a valuable tool from both a safety and innovation perspective. Governments can evaluate impact, secure societal vulnerabilities, develop domestic policy, and track and manage international dynamics. AI companies can mitigate immediate risk, and assist in developing longer-term countermeasures. Compute providers can provide additional security, maintain records and verify compliance, and provide an enforcement mechanism. Other private-sector actors can take a broader set of actions that vary by risk domain.

We also provide two worked examples of what CDDC might look like in practice. The first describes a cyber offense capability identified in a model pre-deployment, and shows how the CDDC process could allow the US government and AI developers to collaboratively leverage that capability for defense in critical infrastructure. The second describes improvements to an open-source biodesign tool that elicits a meaningful boost in pathogen design, and shows how CDDC could allow governments to track progress in this domain and gradually develop countermeasures.

minimum required information-sharing for defenders to mount a response. If implemented, there are likely to be additional bespoke pathways, relationships, and feedback loops.

Finally, we conclude with the following set of recommendations:

*Actions for US government*

1. Congress should assign a coordinator within the US government[3] to receive and distribute reports on dual-use AI capabilities ("DUC reports"), and to develop legal clarity and infrastructure to facilitate reporting from outside government. This should be paired with strengthened reporting requirements for DUCs.
2. Either the President via Executive Order or Congress via legislation should assign agency leads for working groups of "defender" agencies—agencies that receive DUC reports from the coordinator and act on them.
3. Congress should fund the US AI Safety Institute to build capacity for wider government involvement in model evaluations (by enabling agencies to directly perform evaluations, or audit or otherwise be involved in company-run evaluations).
4. National Institute of Standards and Technology (NIST) or alternatively, a non-governmental organization such as Carnegie Mellon University's Software Engineering Institute (CMU SEI) or the Frontier Model Forum (FMF) should lead efforts with AI developers, relevant agencies, and third parties to develop common language for DUC reporting and triage—ideally a "Stakeholder-Specific Vulnerability Categorization" (SSVC) for DUCs.

*Actions for dual-use foundation model (DUFM) developers*

5. DUFM developers should establish clear policies and intake procedures for independent researchers reporting dual-use capabilities, based on Vulnerability Reporting Policies.
6. DUFM developers should create and maintain incident response plans for DUCs and build working relationships with defenders in government, other AI companies, and other relevant non-governmental organizations.
7. DUFM developers should collaborate with working groups (once such groups are developed) to identify capabilities that could help defenders, which can be shared via the CDDC infrastructure.

[3] We make several recommendations on where this coordinator should be housed further in the report.

# Table of Contents

# Introduction: The Necessity of Reporting for Dual-Use AI Capabilities

## What are dual-use AI capabilities?

At the time of writing, there is no standard definition for what constitutes a “dual-use” AI capability, and the term “dual-use” has conflicting definitions in non-AI domains.[4] However, we will draw from the description of capabilities of “Dual-Use Foundation Models” provided in Executive Order 14110 on Safe, Secure, and Trustworthy Development of Artificial Intelligence (The White House, 2023b, sec. 3(k)) which is as follows:

> “[...] high levels of performance at tasks that pose a serious risk to security, national economic security, national public health or safety, or any combination of those matters, such as by:
>
> (i) substantially lowering the barrier of entry for non-experts to design, synthesize, acquire, or use chemical, biological, radiological, or nuclear (CBRN) weapons;
>
> (ii) enabling powerful offensive cyber operations through automated vulnerability discovery and exploitation against a wide range of potential targets of cyber attacks; or
>
> (iii) permitting the evasion of human control or oversight through means of deception or obfuscation.”

[4] The term “dual-use” has sometimes been used to define “items that have both commercial and military or proliferation applications” (Code of Federal Regulations, 2024b), while other times has been used to refer to technologies that “can be used for both beneficial and harmful purposes” (Kulp et al., 2024, p. 4).

### Expert Opinion: We need better definitions

Many experts that we interviewed for this project independently raised the point that current definitions for dual-use capabilities, or what other literature has called *dangerous capabilities*, are too vague—this stood out to us as one of the more robust pieces of community consensus.

We believe that a more precise definition of dual-use capabilities could provide a slew of benefits across the AI governance ecosystem:

1. Clarity for policymakers seeking to understand the possible space of risks posed by AI systems, and to determine which policy actions to take.
2. Clarity for AI companies on what evaluations or model capabilities are within-scope for voluntary or mandatory reporting. An absence of clarity could lead to issues including over-reporting, as companies will want to avoid liability for non-reporting (which may inundate coordinators and defenders)[5], as well as under-reporting, if companies fear that certain reports will fall outside of legal protections or safe harbors related to reporting.
3. Clarity across the ecosystem on how to conceptualize, measure, manage, and report less well-specified capabilities, such as model autonomy.
4. Clarity on establishing thresholds to trigger requirements for AI developers.[6] [7]

We believe that future definitions of dual-use capabilities could be improved, and would like to see future definitions tackle some of the following considerations:

---

[5] For example, see recent issues with CyberTipline managing child sexual abuse material (CSAM) reports (Grossman et al., 2024).

[6] As discussed in Existing infrastructure and gaps in AI governance, policymakers are grappling with the question of how to appropriately define which AI systems to regulate in order to provide public safety, while simultaneously avoiding regulating systems that pose minimal risk. Current approaches rely on a compute-based threshold.

[7] We expect that CDDC should *first* prioritize the most advanced models because we expect them to exhibit the most powerful novel capabilities. However, capabilities available to only the most advanced models today may someday be achievable by models behind the frontier (Pilz et al., 2024), and/or may be elicited through the use of scaffolding or other post-training enhancements (Davidson et al., 2023). This may require the system to collect an expanded set of finders and reported information over time.

1. The possibility that dual-use capabilities may have defensive uses, and may be applied to defend against serious risks in a broad set of domains.
2. Expanding the list of capabilities of concern (with the understanding that the list may be updated over time), including (but not limited to):[8]
    a. Research, development, & acquisition of chemical, biological, radiological, and nuclear (CBRN) weapons
    b. Advanced cyber offensive capabilities
    c. Social engineering capabilities, such as deception, persuasion, manipulation, and political strategy
    d. Model autonomy, including long-term planning, situational awareness, and autonomous replication and adaptation (ARA)
3. Providing specificity on thresholds for which capabilities pose "serious risk" (this will require investment in evaluation science and risk modeling).
4. There may also be non-capability aspects of models that are worth sharing if they are sufficiently worrisome, such as propensity to cause extreme harm.

## Why do dual-use capabilities matter?

Dual-use capabilities (DUCs) pose an urgent and serious risk to security and public safety, because (1) they are likely to emerge from ongoing AI development trends, (2) they can be misused by actors who have or gain access to the model, and (3) autonomous AI systems may themselves pose significant risks. There are also additional benefits to establishing better information-sharing and coordination around DUCs, including:

- Providing insight into how capabilities of AI systems more broadly will develop in the future.
- Alerting defense-relevant actors to defensive uses of dual-use capabilities.
- Strengthening visibility of AI capabilities to support better-informed AI governance.

[8] This list includes capabilities that either (a) have been subject to evaluation at advanced AI companies, such as autonomous replication and adaptation (Kinniment et al., 2024), (b) have been noted as existing or potential targets of evaluation by advanced AI companies (Anthropic, 2023a; OpenAI, 2023a) or by governments (Department for Science, Innovation & Technology, 2024b; The White House, 2023b) or (c) are discussed in (Shevlane et al., 2023).

## Dual-use capability risks

DUCs are likely to emerge at some point if current AI development trends continue; to some extent, recent research has suggested that certain dual-use capabilities are already starting to emerge from cutting-edge models. For instance, teams of AI agents have been found to be capable of exploiting zero-day vulnerabilities in some real-world systems (Fang et al., 2024). Recent AI progress has been rapid, driven by algorithmic progress, investment, and training compute, among other factors (Epoch AI, 2024). In just the past few years, advanced AI has surpassed human capabilities across a range of tasks. These include image classification, basic reading comprehension, and natural language inference (see Figure 1). AI systems are able to handle increasingly complex tasks: they can write fluently and at length, write code, summarize complex documents, and generate increasingly realistic videos from natural language prompts.

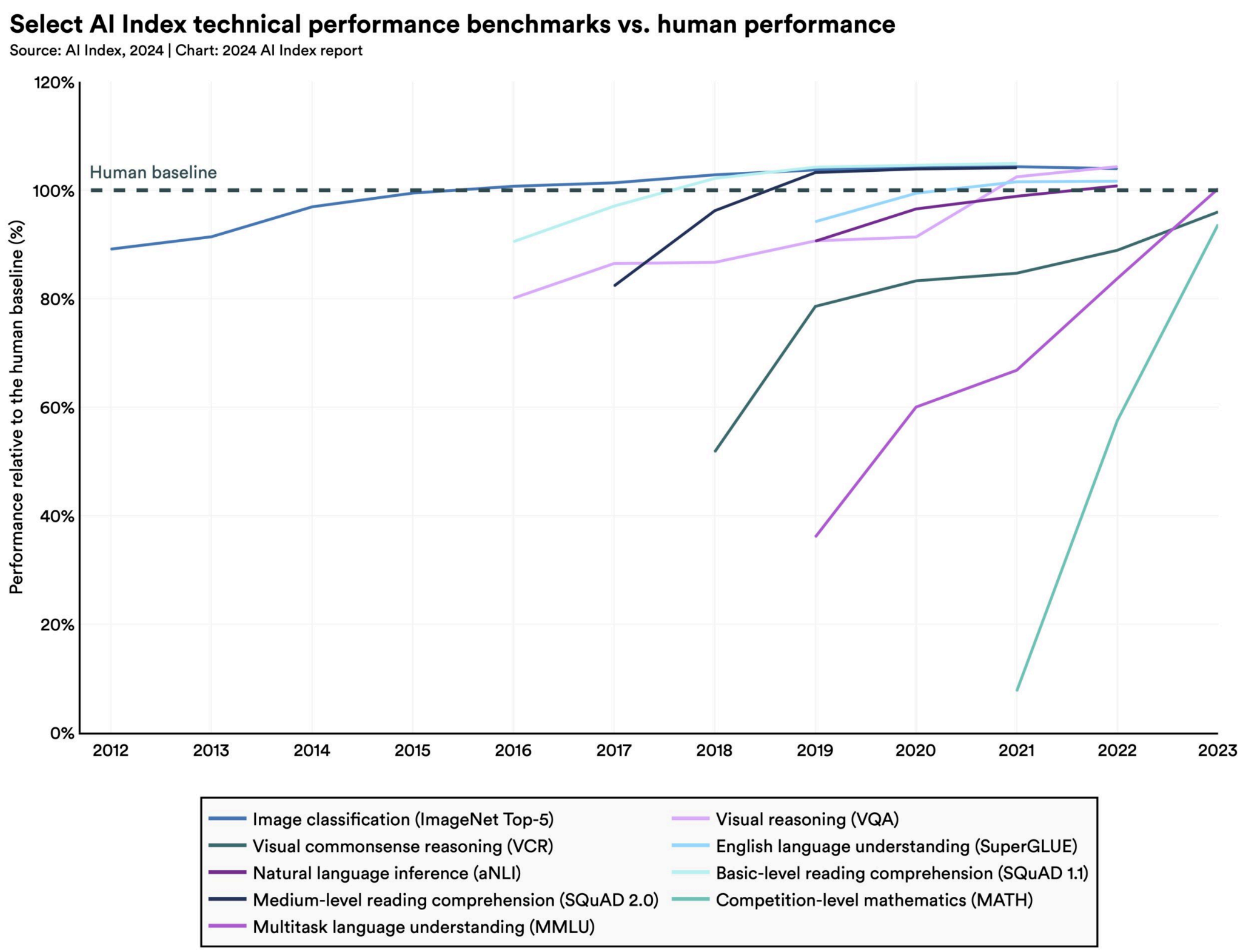

*Figure 1 (from Perrault & Clark, 2024).*[9] *AI systems have surpassed human performance on multiple benchmarks.*

[9] Graphic licensed under [CC BY-ND 4.0] (https://creativecommons.org/licenses/by-nd/4.0/).

Rapid progress in AI systems will likely continue in the future. Model performance on training objectives has scaled predictably with increased compute used in training. Leading model developers like OpenAI and Google Deepmind have plans to spend tens of billions of dollars to train more powerful models (Seal, 2024). Hardware manufacturers are expanding the production of specialized AI chips, suggesting additional resources will be available to train more sophisticated models (Chowdhury, 2023). As models gain more sophisticated capabilities, some of these will likely qualify as dual-use under our definition above. Furthermore, new AI capabilities can emerge unexpectedly and remain undetected for some time (Anderljung, Barnhart, et al., 2023).

**DUCs can be misused by actors who have or gain access to the model.** Harms from misuse of current AI models are already being felt, including from automation of fraud, non-consensual pornographic image generation, and the use of image recognition systems to identify and persecute minority populations (Anderljung & Hazell, 2023). As AI systems are developed with increasingly advanced capabilities, these could be weaponized by malicious actors in a variety of domains, to "perform cyberattacks, run disinformation campaigns, and design biological or chemical weapons, [and] almost certainly continue to lower the barriers to entry for less sophisticated threat actors" (Department for Science, Innovation & Technology, 2023). While model misuse is largely gated on model access, uncontrolled access can be gained either through the open-sourcing of model weights, or by theft of model weights. This suggests that, while AI companies, governments, and compute providers can all play a role in mediating model access, and thereby reduce misuse, there is still a need to prepare other proactive defenses. CDDC allows for this preparation to be more efficient and comprehensive.

**Autonomous AI systems may themselves pose significant risks.** Some researchers have raised the issue that sufficiently advanced AI systems may have the capacity and/or propensity to engage in significantly harmful behaviors that their users do not intend, including power-seeking behaviors such as self-exfiltration, preventing shutdown, and controlling their environment (Bengio et al., 2024; Harris et al., 2024; Hendrycks et al., 2023). While loss-of-control risk from AI is novel, loss-of-control risks from technology developments are not unprecedented. In the biology domain, some engineered pathogens are carefully guarded in biosafety labs, because a breach of containment could mean spreading serious illness to outside populations (Bloom & Lentzos, 2024); in the cyber domain, Stuxnet is a prime example of a computer worm that was more effective than expected, and infected computer systems much more widely than its creators intended (Zetter, 2014).

## Benefits of reporting dual-use capabilities

**DUCs of advanced AI systems provide insight into how capabilities of AI systems more broadly will develop in the future.** Due to cumulative increases in algorithmic efficiency and hardware price

performance, training yesterday's most powerful AI models is possible today for a much lower investment (Ho et al., 2024; Pilz et al., 2024), and thereby available to a much wider set of actors. If this trend continues, we should expect the capabilities of today's powerful models to be replicable in the future for a much lower cost–and this lowers the barrier for malicious or incautious developers to access powerful capabilities simply by developing a model themselves. However, large compute investors may maintain an edge by benefiting from efficiency improvements in addition to smaller actors (though the gap between them may shrink over time if certain model capabilities reach diminishing returns on performance (Pilz et al., 2024). Tracking the most advanced DUCs today will provide defenders with more time to develop plans for a future with more widespread AI capabilities, or to take actions to limit certain capabilities from being developed or diffused in cases where defenses are insufficient or impossible to develop.

**Some DUCs may have defensive uses, which could aid in mitigating serious risks.** For example, the ability to identify zero-day cyber exploits to cripple critical infrastructure at scale is synonymous with the ability to identify zero-day exploits in order to allow defenders to *patch them* at scale. Similarly, certain advances in AI-bio design tools may improve our ability to develop vaccines (Coalition for Epidemic Preparedness Innovations, 2023). While not all DUCs will have this property, it is important to note that the CDDC infrastructure can also be used to give defenders early warning and potentially access to defensive uses of AI.

**Reporting DUCs can strengthen visibility on AI capabilities to support better-informed AI governance.** As stated in Kolt et al. (2024), "information is the lifeblood of good governance." Providing relevant US government employees and policymakers with up-to-date information on AI capabilities can facilitate better-informed responses, for both short-term emergency response and longer-term policy response.

**To that point, one critical question when discussing reporting for DUCs is: *What can defenders do when given information about dual-use capabilities?*** While our process focuses primarily on reporting information, in order to provide value, reporting must lead to concrete actions when necessary. Below, we provide a set of actions that governments, AI developers, compute providers, and other private-sector actors could take to respond to information about DUCs.[10]

---

[10] While not focused on dual-use capabilities explicitly, (Department of Homeland Security, 2024, pp. 16–17) provides a list of potential mitigations to attacks on critical infrastructure using AI, which could be used as a supplementary resource to the following table.

## Table 1: Potential responses to dual-use capabilities

| Responses | Examples |
|---|---|
| **Actions by governments** | |
| Evaluate impact | Perform research into the implications of the capability on critical infrastructure, for instance whether the electricity grid computer systems require hardening to potential increased frequency of AI-powered cyberattacks, or future quantum cyberattacks (Tierney, 2024). |
| | Order evaluations of models on a similar training scale, or of other models using similar post-training enhancements, to determine whether the DUC is more widespread. In other critical infrastructure domains, comparisons to existing similar systems are often preferred over completely new safety evaluations (European Union Agency for Railways, n.d.). |
| Secure societal vulnerabilities to the capability | Use new capabilities defensively: such as by using the model capability to harden cyber infrastructure (Venables & Hansen, 2024). |
| | Request or compel external actors to guard against misuse—e.g., request DNA synthesis companies develop advanced screening techniques to prevent users from synthesizing dangerous novel pathogens (Administration for Strategic Preparedness and Response, 2023). |
| Develop domestic policy | Request or compel the model owner(s) to bolster security around the model against theft by adversaries (Nevo et al., 2024). |
| | Pass legislation for a higher-risk technology environment than we have today, such as standing up an oversight entity for AI development and deployment (Romney et al., 2024). |
| Track and manage international dynamics | If governments are concerned about foreign actors using a capability, they can gather information via the intelligence community, as has been the case for monitoring nuclear weapons proliferation (Congressional Research Service, 2023). |
| | Pass relevant information to other governments that might be affected so they can prepare adequately. For instance, information could be shared with trusted intelligence partners in the Five Eyes group, with specific countries bilaterally, or with broader non-proliferation international mechanisms such as the Australia Group (Department of Foreign Affairs and Trade, 2023). |
| | Confidence-building mechanisms where the USG provides a credible signal that it doesn't intend to weaponise that capability against others and intends to act responsibly (Imbrie et al., 2023). For instance, parties to the biological weapons convention are obliged to submit annual reports to other states parties providing evidence and assurance of the peaceful nature of any advanced biotechnology facilities (Office for Disarmament Affairs, n.d.). |
| | Collaborate internationally to develop a governance framework for the capability, as has been done for past dual-use technologies, such as via the Chemical Weapons Convention (Organisation for the Prohibition of Chemical Weapons, 2024). |
| | Institute additional export controls to slow malicious foreign actors from accessing or developing the model capability, building on existing BIS semiconductor export controls (Bureau of Industry and Security, 2023). |
| **Actions by dual-use foundation model developers** | |
| Mitigate immediate risk | Implement access restrictions on deployed models that may present risk, such as by user restrictions, access frequency restrictions, capability restrictions, use case restrictions, or decommissioning (O'Brien et al., 2023). |
| | Pause development or deployment of the model in question until proper countermeasures have been developed (Alaga & Schuett, 2023). |
| | Secure model weights and other important IP to prevent model theft and misuse (Nevo et al., 2024). |
| Help develop countermeasures | Aid governments with top-tier AI expertise to develop countermeasures, similarly to how in the biosecurity domain USG set up the National Science Advisory Board for Biosecurity to get expert advice on relevant government decisions (Office of Science Policy, 2024). |
| | Assist governments to use the model in question to identify critical vulnerabilities or other risks, such as cyber vulnerabilities or particular worst-case uses of the model capability. For instance, the UK AISI is building up capacity to evaluate advanced AI models (Department for Science, Innovation & Technology, 2024b). |
| **Actions by compute providers** | |
| Provide additional security | Securing against leaks or theft of critical IP, such as model weights (Heim et al., 2024). |
| Record keeping and verification | Monitoring for suspected violations of reporting requirements, akin to how financial institutions are legally obliged to do anti-money-laundering know your customer (KYC) checks (Dow Jones, 2024). |
| Enforcement | Refusal to host or deploy a model (Heim et al., 2024). |
| | Disabling AI systems that display unwanted activity, such as a cyber worm-like-system (Heim et al., 2024). |
| **Actions by other private-sector actors** | |
| Varies by risk domain. Illustrative examples by domain include: | Biology: Researchers and gene synthesis companies could collaborate to create gene screening techniques that are not based solely on sequence similarity to known pathogens, to reduce risk of gene synthesis production of novel pathogens (Balaji et al., 2022). |
| | Cybersecurity: Cybersecurity companies could collaborate with AI developers to use a model with advanced cyber offensive capabilities to identify vulnerabilities in critical infrastructure and patch them (Tierney, 2024). |

The goal of the action set described above is to prevent significant harm through use of AI capabilities. However, success relies on at least one of two preconditions being met, to prevent exploitation of the capability before countermeasures are developed:[11]

[11] There is a third precondition, but only related to loss-of-control: sufficient security. If there is loss-of-control risk, this precondition requires that sufficient security measures are in place to maintain control of a model/system/agent.

- **Limited external awareness:** Malicious actors or adversaries are not yet aware of the availability of the capability
- **Limited unwanted access:** If such actors *are* aware of the capability, they do not have access to an AI system that possesses this capability

Both of these factors increase the time available to defenders. *External awareness* about a capability can arise in a number of ways, including through intentional or unintentional information leaks, espionage, and public displays of capability identification (such as a Twitter post) of a deployed model. *Unwanted access* to a capability can arise in the near-term from the theft and copying of model weights, by premature open-sourcing of model weights, or poorly-scoped internal permissions to use a model (such as by an untrustworthy employee at an AI company), and in the longer term by additional developers training equivalent or more powerful models, as techniques and compute to cheaply train powerful AI systems proliferate.

**Therefore, the goal of CDDC is to provide defenders with maximal time and the necessary information to respond**—primarily by maintaining a carefully-sequenced and secure information flow to ensure that awareness of the capability is limited to the necessary actors.[12]

## What are the gaps in the current ecosystem?

**At a high level, we find several notable gaps in the existing AI governance infrastructure that limit the use and efficacy of capabilities reporting, including:**

1. An *evaluation gap*: there is a lack of a robust evaluation system, including both (a) technical evaluations, and (b) agreement on what capabilities pose dangers to public safety and/or security.[13]

2. A *reporting gap*: several disincentives to reporting information to government encourage non-reporting, across multiple types of finders.

3. A *coordination gap*: we find challenges with the existing mandatory capabilities reporting system in the US established under Executive Order 14110 Sec. 4.2(a)(i) that limit the US federal government's potential use of capabilities information.

---

[12] This approach is inspired in part by Coordinated Vulnerability Disclosure (CVD), a process in cybersecurity used to accomplish similar goals (Householder et al., 2017).

[13] In this paper we do not attempt to solve this evaluation gap; we take as given that evaluation techniques exist and will be improved upon.

4. A *defense gap:* there is a lack of clear ownership for specific risk areas, especially for AI-specific emerging risks (such as model autonomy or deception).

For a more in-depth description of the existing infrastructure and gaps in AI governance as related to CDDC, see below.

## Methods

This project draws on a combination of desk research and semi-structured interviews. Desk research was used to preliminarily identify gaps in the disclosure landscape and to come up with tentative recommendations addressing said gaps. We then workshopped those initial findings via over 60 semi-structured interviews with experts from government, industry, civil society, and academia.

The semi-structured interviews were by far the lengthiest and most important phase of our research. For the first set of gaps that we identified in the disclosure landscape (on finders—see below), we piloted a workshop format (with multiple interviewees in one group). However, after that pilot, we settled on single-person interviews for the remaining bulk of our research, to maximize our reach among busy policymakers and industry figures.

We interviewed 66 experts in total.[14] The interviews focused on a range of topics, from single policy recommendations to overarching considerations regarding the structure of our disclosure ecosystem. Interviewees included personnel from various governments, staff members at companies developing and deploying AI, and experts from think tanks, third-party AI evaluators, and academia. Participation in this research does not necessarily imply endorsement of this report or its findings, and the views expressed by individuals involved do not necessarily reflect those of their respective organizations.

We iterated the design of our CDDC framework as we went, based on new recommendations from interviewees and their critiques of our existing plans. As a result, our final list of gaps and recommendations differs substantially from the findings of our initial research.

[14] Including the initial workshop (with 7 attendees).

# Proposal: Coordinated disclosure of dual-use capabilities

As discussed above, a secure, clear reporting system for information regarding dual-use AI capabilities (DUCs) could enable society to mitigate the risks they pose. Evidence of the existence of specific capabilities can guide responses from defenders including model developers, government agencies, policymakers, and other private sector actors.[15]

We propose Coordinated Disclosure of Dual-Use Capabilities (CDDC) as a process to guide information-sharing about DUCs, based loosely on the concept of coordinated vulnerability disclosure (CVD) in cybersecurity.[16] CDDC describes a set of reporting pathways from finders, through a coordinator, to defenders, optimized to allow defenders maximal time to respond to the existence of a given capability before information about the capability is made public (Map 1). The process centers around an information clearinghouse (the "*coordinator*") which receives evidence of DUCs from *finders* via mandatory and/or voluntary reporting pathways, and passes noteworthy reports to *defenders* for follow-up (i.e., further analysis and response).[17] This aims to provide the government and dual-use foundation model (DUFM) developers with a comprehensive overview of AI capabilities that could significantly impact public safety and security, and give them time to respond and implement countermeasures by providing early and private access to information about DUCs.[18]

[15] US Representative Ted Lieu highlighted the value of such evidence in a recent POLITICO interview: "If you just say, 'We're scared of frontier models' — okay, maybe we should be scared [...] But I would need something beyond that to do legislation. I would need to know what is the threat or the harm that we're trying to stop" (Bordelon, 2024).

[16] The potential application of lessons from CVD to the AI domain has also been explored in (Cattell, Ghosh, & Kaffee, 2024), which focuses on the disclosure of unintended model behavior.

[17] In more detail: DUC finders report evidence of a DUC to the model developer and to a legally protected coordinator in civil society or (ideally) in government; this coordinator triages DUC reports and assigns them to a government agency (and possibly other evaluators) for follow-up, including convening working groups for serious DUCs.

[18] E.g., they will need to establish the potential impact of the advanced capability, verify peer companies' models do not pose potential harms, and adopt/adapt defensive techniques in the public and private sector.

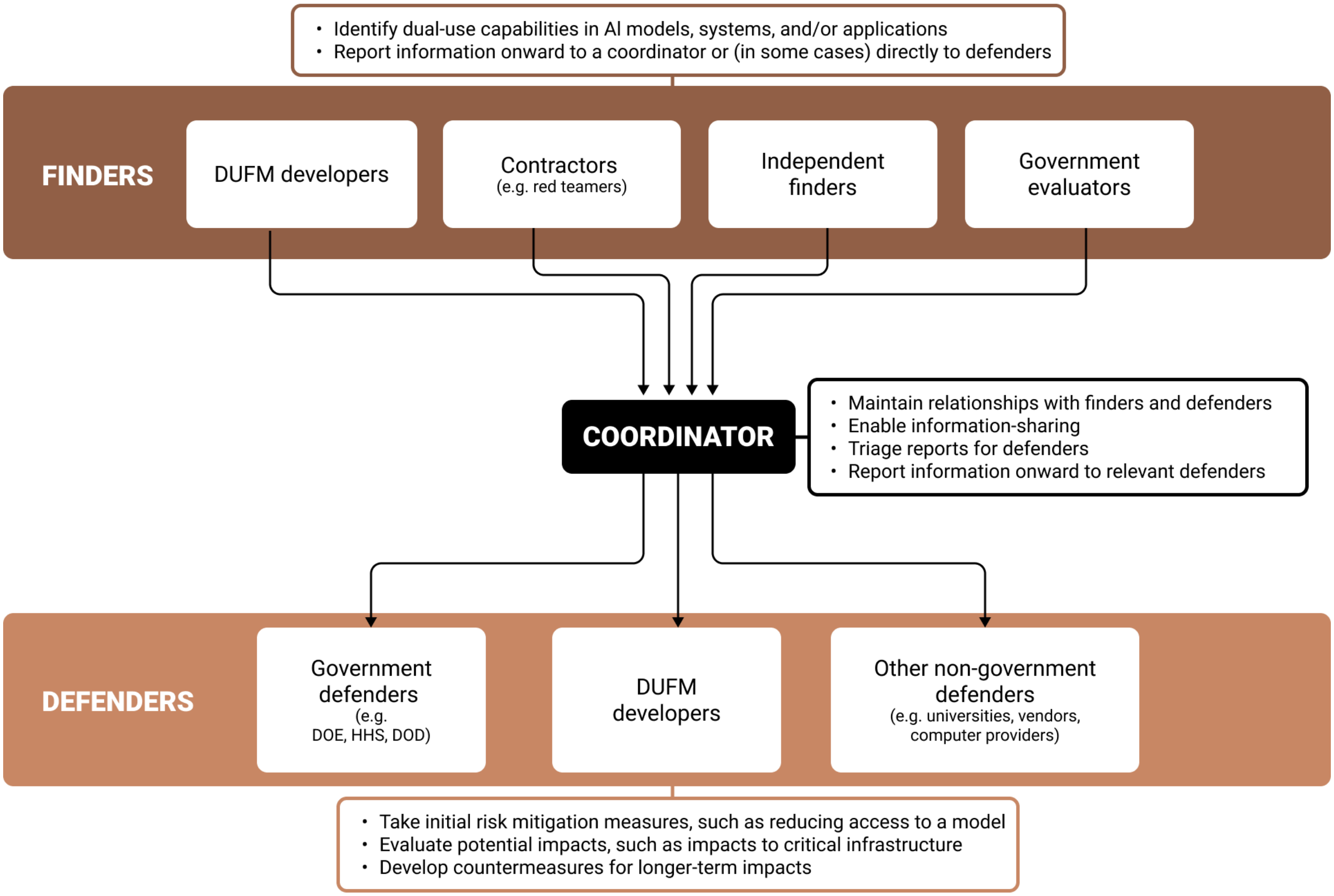

*Map 1: Basic Structure of CDDC*[19]

## Roles within CDDC

The three primary sets of roles within a coordinated disclosure ecosystem are finders, coordinators, and defenders (see Map 1 for a visualization of their relationships).

[19] Importantly, the suggested structure is not designed to preclude other forms of information-sharing (such as bespoke arrangements between AI companies and specific federal agencies, company-to-company information sharing, or alternative paths for whistleblowers). We primarily hope for the establishment of a central coordination point as a floor (rather than a ceiling) to ensure the minimum required information-sharing for defenders to mount a response. If implemented, there are likely to be significant additional pathways, feedback loops, and bespoke relationships. Map 3 (Recommendations) is one step closer to implementation-level detail.

**Finders are parties that initially identify dual-use capabilities.** This is a broad set of actors, which includes:

- **Dual-use foundation model (DUFM) developers themselves**, as developers may have in-house teams conducting dual-use capability evaluations.
- **Contracted evaluators** for one or more advanced AI developers, such as the Model Evaluation & Threat Research organization (METR), which has previously partnered with OpenAI and Anthropic to test their models.
- **Independent security researchers**, who may identify dual-use capabilities as experts engaged by advanced AI developers during pre-deployment testing[20] and after deployment, or as unaffiliated researchers conducting testing on advanced AI models after deployment.[21]
- **Evaluators within governments**, to the extent that such capacity is developed, such as (within the US) the US AI Safety Institute[22]; Department of Commerce (DOC) via the National Institute of Standards and Technology (NIST)[23]; The Department of Energy (DOE)[24];

---

[20] E.g., as part of the OpenAI Red Teaming Network (OpenAI, 2023c).

[21] E.g., similar to the CMU and CAIS researchers who identified a class of broadly applicable adversarial attacks on LLMs in August 2023 (Knight, 2023).

[22] USAISI may launch projects that involve testing and evaluating advanced AI models, systems, and agents, including at the pre-deployment stage (NIST, 2024).

[23] EO 14110, 4.1(a)(i)(C) tasks NIST with "launching an initiative to create guidance and benchmarks for evaluating and auditing AI capabilities, with a focus on capabilities through which AI could cause harm, such as in the areas of cybersecurity and biosecurity."

[24] EO 14110 Sec. 4.1(b) tasks DOE to "develop and [...] implement a plan for developing the Department of Energy's AI model evaluation tools and AI testbeds. The Secretary shall undertake this work using existing solutions where possible, and shall develop these tools and AI testbeds to be capable of assessing near-term extrapolations of AI systems' capabilities. At a minimum, the Secretary shall develop tools to evaluate AI capabilities to generate outputs that may represent nuclear, nonproliferation, biological, chemical, critical infrastructure, and energy-security threats or hazards."

and the Department of Homeland Security (DHS)[25]; within the UK, this may include the UK AI Safety Institute (UKAISI).[26]

The *coordinator* in this process acts as an information clearinghouse, and is responsible for providing infrastructure for collecting, triaging, and passing along reports to defenders.[27] The coordinator ideally has relevant competencies and high-trust relationships with both finders and defenders in order to effectively source and direct key info about DUCs, and is resourced appropriately. While we recommend a coordinator situated within the US government (Recommendation 1), a non-governmental coordinator may be possible.

*Defenders* are parties that are capable of taking steps to help mitigate the harms of a DUC. While the set of relevant defenders is likely to vary depending on the particular DUC,[28] we identify three major defender categories:

- Risk-specific government agencies that can (for example) patch vulnerabilities to AI-assisted attacks on critical infrastructure.
- Advanced AI companies (the frontline developers and often deployers of advanced AI systems), who may possess unique expertise, access, and legal permissions regarding the AI system in question.
- Other non-governmental defender organizations, such as compute providers, universities, cyber defense teams, and other sector-specific actors who are responsible for employing appropriate defenses in their domains, such as DNA synthesis companies.

---

[25] EO 14110, Sec. 4.4(a)(i) tasks DHS to collaborate with OSTP and DOE to "evaluate the potential for AI to be misused to enable the development or production of CBRN threats, while also considering the benefits and application of AI to counter these threats [...] (A) consult with experts in AI and CBRN issues from the Department of Energy, private AI laboratories, academia, and third-party model evaluators, as appropriate, to evaluate AI model capabilities to present CBRN threats."

[26] One of UKAISI's three core functions is to "develop and conduct evaluations on advanced AI systems" (Department for Science, Innovation & Technology, 2024b).

[27] For more information on the coordinator role, see Recommendation 1.

[28] For example, for bioweapons development, the Department of Health and Human Services could be a key defender, while for a loss-of control risk, the model developer may be the most relevant party involved in primary response, by taking swift action to contain or decommission the model. See the case study for two distinct stories that explore how CDDC could play out with DUCs in different domains.

## Hypothetical Scenario: Cyber Capability

*The following scenario gives one example of how a potential dual-use capability could arise, and how the CDDC framework could be applied to mitigate negative outcomes.*

**The capability:** In this scenario, a leading AI company develops a model that exhibits a leap in capability to autonomously detect, identify, and exploit cybersecurity vulnerabilities in critical infrastructure, which arises prior to model deployment.

**The risk:** A state or non-state actor could get a hold of this capability or independently develop a similar capability and use it to attack critical infrastructure throughout the US and other nations.

**The finder:** The AI developer identifies this during evaluation partway through a training run.

**The reporting process:** The company reports directly to the government coordinator (Rec. 1).

**The triaging process:** The coordinator passes information along to CISA (DHS), the working group lead for cyber threats including to critical infrastructure (Rec. 4).

**Immediate response:** CISA verifies the evaluation using capacity developed via Rec. 3, and finds the report alarming. They escalate the information to the NSC. Meanwhile, the AI developer bolsters security around the model and institutes deployment mitigations (Dragan et al., 2024).

**Longer-term response:** US federal agencies, including Sector Risk Management Agencies (SRMAs), collaborate with the AI developer to use the model to identify vulnerabilities across multiple critical infrastructure sectors for patching (Rec. 7); US policymakers impose security requirements on domestic AI developers to counter potential theft by adversaries.

**Epilogue:** The success rate of cyberattacks on critical infrastructure *in general* in the US is significantly reduced over the course of the next year, because major exploits have been identified and patched. One year later, a US adversary develops a model with a similar level of offensive cyber capabilities. When they attempt to use the model to compromise US critical infrastructure, the effect is minimal.

# Existing infrastructure and gaps in AI governance

Before discussing our recommendations, we will explore the current status of AI governance as it relates to CDDC, and ask: *What building blocks do we have to work with? What else is needed?* In short, the current AI governance ecosystem features many elements of CDDC, but is nevertheless missing critical parts of a well-functioning reporting system. The major issues our research identified regarding the current state of AI governance are outlined in the map below:

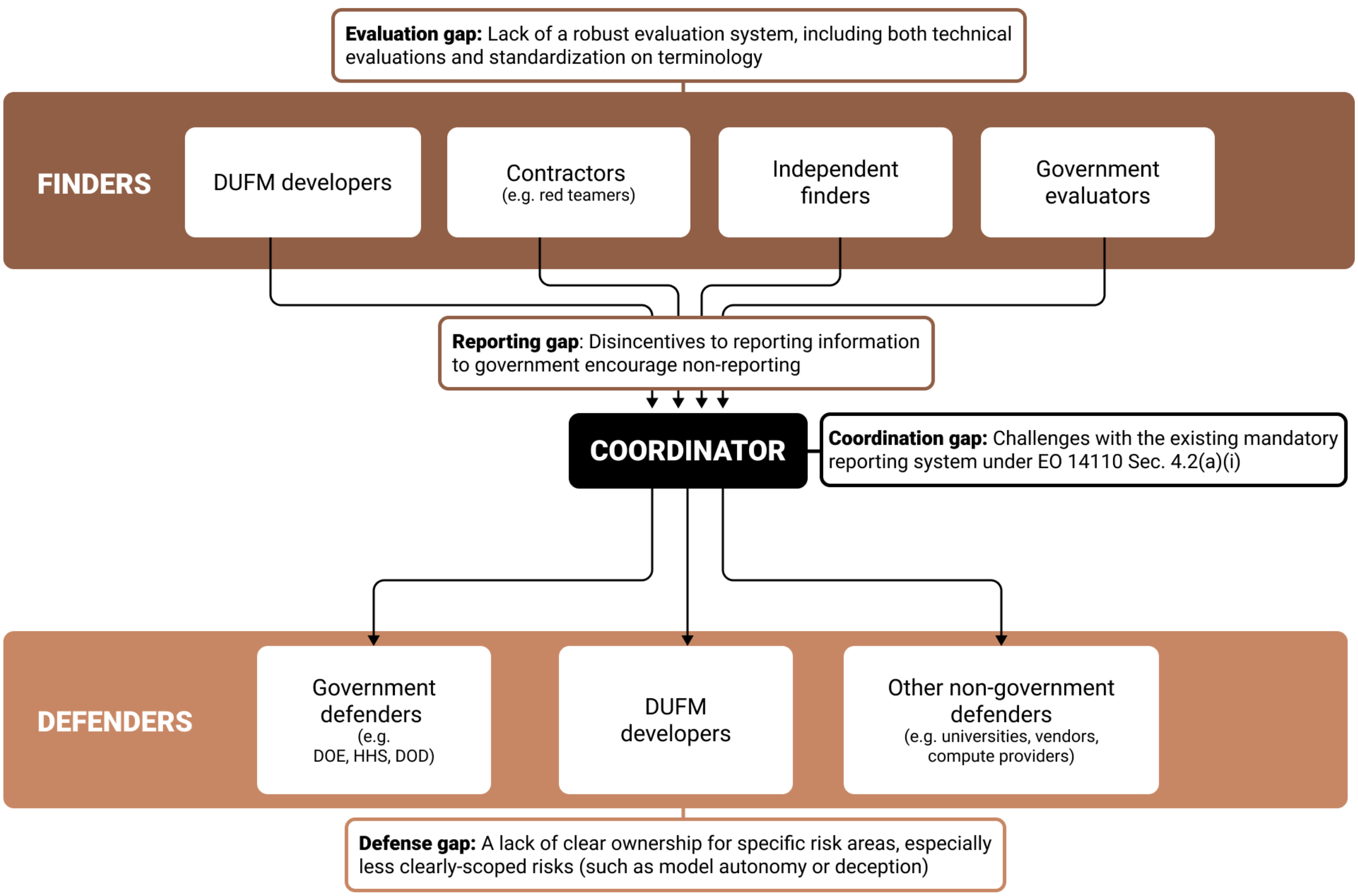

*Map 2: Gaps in existing governance related to CDDC*

This section reviews current laws and practices as they relate to different actors in CDDC, and notes significant remaining gaps that our recommendations aim to address.

## Finders

### Policies and practices

Red-teaming and evaluation to elicit capabilities of advanced AI models has emerged as a potentially promising tool for informing society's approach to AI governance and risk management in the past few years.[29] [30] Multiple types of evaluators, reflective of the different sets of finders in our framework, are at different stages in their capacity to evaluate AI systems in practice.

AI companies are already evaluating models prior to deployment to check for dual-use capabilities, such as the capacity to enhance cyber operations, enable the development of biological or chemical weapons, and autonomously replicate and spread (Anthropic, 2023a; OpenAI, 2023a). Third-party evaluators have been involved in pre-deployment evaluation and red teaming, including both individual red-teamers and specialized teams (Anthropic, 2023a; METR, 2024; OpenAI, 2023c).

Meanwhile, the US and UK governments in particular have taken a number of steps to improve government capacity for model evaluation. In the US, President Biden's Executive Order 14110 on "Safe, Secure, and Trustworthy Development and Use of Artificial Intelligence" directs multiple federal agencies to take steps to develop US capacity for AI evaluation, including the development of AI testbeds, and consultation with experts to evaluate model capabilities to present CBRN threats (The White House, 2023b). The US, UK, and other countries have stood up internal expert groups in the form of AI Safety Institutes,[31] and the US and UK Institutes have established a memorandum of understanding for collaboration on the development of evaluations for advanced AI (U.S. Department of Commerce, 2024). In 2023, the UK government welcomed commitments from AI companies in attendance at the Bletchley Park AI Safety Summit to "work with the Institute to allow for pre-deployment testing of their frontier AI models and commitments to work in partnership with other countries' Institutes including the US." Following that, the 2024 AI Seoul Summit resulted in AI safety commitments from major technology companies including risk assessment of model capabilities across the AI lifecycle (Department for Science, Innovation & Technology, 2024c), as well commitments from twenty-seven nations and the European Union to develop thresholds for severe risks and collaborate on AI safety guidelines (Department for Science, Innovation & Technology, 2024d).

---

[29] For example, this approach has been suggested in the 2024 Senate AI Policy Roadmap: "[...] the AI Working Group encourages the relevant committees to consider a resilient risk regime that focuses on the capabilities of AI systems [...] the risk regime should tie governance efforts to the latest available research on AI capabilities and allow for regular updates in response to changes in the AI landscape" (Schumer et al., 2024).

[30] For an overview of relevant terms in AI evaluations, see A starter guide for evaluations (Hobbhahn, 2024).

Additionally, the public[32] has played a role in evaluating models after deployment, with results often being published in academic journals, pre-print servers, or news or social media (Buolamwini & Gebru, 2018; Roose, 2023a; Yong et al., 2024).

## Gaps

Evaluation gap: Existing capability evaluations, as well as the current status of evaluations in AI governance, may be insufficient to ensure that capabilities of advanced AI models are tracked reliably.

Several of our interviewees shared concerns that existing dual-use, or "dangerous capability evaluations" (as described in Shevlane et al., 2023) are insufficient to capture the risks they worry may emerge from advanced AI models. Additionally, interviewees raised the challenge that existing evaluations are not interoperable—in some cases, an evaluation for Model A at Company A cannot be easily translated for use on Model B at Company B. A more detailed technical critique of current evaluation methods can be found in (Anwar et al., 2024, pp. 16-19, 50-53). Furthermore, there are not yet public standards on what evaluations should be run, outside of high-level guidance from

[31] USAISI's initial projects include: "advancing research and measurement science for AI safety, conducting safety evaluations of models and systems, and developing guidelines for evaluations and risk mitigations" (NIST, 2023b). UKAISI, meanwhile, plans to "[...] focus on the most advanced current AI capabilities and any future developments, aiming to ensure that the UK and the world are not caught off guard by progress at the frontier of AI", and undertake three major functions: "Develop and conduct evaluations on advanced AI systems [...] Drive foundational AI safety research […] Facilitate information exchange" (Department for Science, Innovation & Technology, 2024a). UKAISI recently released an open-source evaluation platform with the goal of accelerating research on AI safety evaluations (Department for Science, Innovation & Technology, 2024b).

[32] Including, for example, academic researchers and independent security researchers.

government.[33] [34]

**How does CDDC help?** While the evaluation gap is much broader than our recommendations aim to address, Recommendation 3 aims to mitigate this issue by improving US government capacity to run or oversee evaluations.

**Other issues include:** Lack of government evaluators/government model access, especially pre-deployment[35]; and limited external research access to models (discussed further in Bucknall & Trager, 2023).[36]

## Reporting from Finders to Coordinators and/or Defenders

### Policies and practices

We divide reporting here into two buckets: reporting from finders to government, and from finders to AI companies.[37] We focus on government and AI companies because they are the best-situated parties to manage an initial response to dual-use capabilities. The primary question here is: How are coordinators/defenders receiving information on dual-use capabilities, and what are the gaps?

---

[33] For example, Executive Order 14110 provides some direction for AI companies evaluating models, by defining risks that dual-use foundation models may present, such as cyber, biological, radiological, and nuclear risks (The White House, 2023b). However, this guidance is high-level and does not include concrete evaluation practices, thresholds, or other specifics.

[34] While technical standardization of evaluations may be difficult to achieve in such a fast-moving space (and, as Anderljung et al. (2023) argue, formulaic approaches to evaluation may actually be harmful), certain targets for standardization may be useful while being easier to produce, such as specific redline capabilities to test for, or what information should be provided in an evaluation report.

[35] POLITICO reports that there are some technical issues to be resolved with regard to access. For example, Anthropic's co-founder Jack Clark has stated that "pre-deployment testing is a nice idea but very difficult to implement," and that Anthropic is working with UKAISI to figure out how to resolve this (Manancourt et al., 2024). However, in a recent interview Secretary of Commerce Gina Raimondo states that the US AI Safety Institute will conduct pre-deployment testing of all new advanced AI models (Henshall, 2024).

[36] It is worth noting that there is a tradeoff between external researcher access and security of information about dual-use capabilities: providing resources for external researchers increases the chance that issues with models are discovered, but also increases the number of actors who could publicly disclose those issues. There may be ways to reduce the tension in this tradeoff, such as providing incentives for finders to securely and privately disclose findings.

[37] There is additional work to be done to identify reporting pathways to other potential defenders, such as compute providers.

As discussed above, AI companies receive information on AI capabilities primarily through running internal evaluations, or by sourcing evaluations from third-party evaluators.

The US government receives information on AI capabilities in a number of ways. First, Executive Order 14110 mandates reporting from companies to the government on certain evaluation results concerning potential dual-use foundation models (The White House, 2023b). Second, there are no prohibitions on voluntary reporting from AI companies or independent researchers to government (however, there is little guidance on how to initiate this reporting). Third, there are initial steps forming for government evaluation of models and intergovernmental information-sharing, such as in the USAISI-UKAISI Memorandum of Understanding (U.S. Department of Commerce, 2024) though what information can be legally shared under that agreement is yet to be seen.

### Gaps

Reporting gap: Disincentives around reporting can lead to nondisclosure of evaluation results to the government.

There are several major barriers for non-governmental parties reporting to government:

- AI companies are concerned about potential leaks of shared information due to poor security outside of their organizations, leading to loss of intellectual property, divulging of trade secrets, and/or potential reputational damage.
- All parties may face regulatory enforcement for producing certain results, such as:
    - If a model is capable of inferring classified information. For example, unauthorized receipt and/or mishandling of nuclear secrets (such as by someone prompting the model) could incur criminal penalties under the Atomic Energy Act and the Espionage Act, though the use of AI to do so may pose novel legal questions.[38]
    - If, in order to verify that a potentially illegal capability exists, the act of verification itself constitutes a crime (such as verifying the successful design of a controlled substance by producing it).
- There are no whistleblower protections for employees or contracted evaluators of AI companies, many of whom have signed non-disclosure agreements.[39]
- There is no clear reporting hub for the government to intake voluntary reports on AI capabilities, including by independent finders and by companies.

[38] Relevant laws: Espionage Act: 18 U.S.C. § 793 (Gathering, transmitting or losing defense information) AEA: 42 U.S.C. § 2274 (Communication of Restricted Data), 2276 (Receipt of Restricted Data), 2277 (Disclosure of Restricted Data).
[39] This issue was recently highlighted by a number of former and current employees at OpenAI and Google DeepMind in a “right to warn” proposal (Right To Warn, 2024).

**How does CDDC help?** In Recommendation 1, we propose that a coordinator within government develop the legal and infrastructural tools to resolve these issues. Recommendation 4 also aims to simplify the reporting challenges through the development of standardized reporting forms. While not focused on government, Recommendation 5 aims to improve reporting from independent finders to AI companies.

**Other issues include:** Insufficient reporting infrastructure and response staffing can mean that finders are unable to securely and effectively relay reports (Pesce, 2024); in the case that contracted evaluators and AI companies disagree on the severity of the finding, there are no clear ways to adjudicate such disagreements; a lack of trusted reporting pathways can also lead finders to opt for *premature public* disclosure, which can incidentally flag capabilities to malicious actors; a lack of specified deadlines for any parties reporting dual-use capabilities.

## Coordinators

### Policies and practices

Within the US government, there is no primary party that plays the role of an information clearinghouse for dual-use capabilities today. As discussed above, there is some information *collection* (such as by the Bureau of Industry and Security (BIS), the main intake body for mandatory reporting of select capabilities of dual-use foundation models as described in Executive Order 14110).

Within industry, there have been significant challenges for company-to-company information sharing of evaluation results, largely due to (1) industry fears of leaked intellectual property, and (2) a lack of clarity regarding whether such sharing would violate antitrust regulation. While the Frontier Model Forum (FMF) has been carving out a role as non-governmental coordinator for advanced AI companies with a goal of providing a forum "for cross-organizational discussions and actions on AI safety and responsibility" related to safety standards and evaluations (Frontier Model Forum, 2024), the extent to which the FMF will provide space for the sharing of evaluation *results* is unclear, and companies that are not FMF members may not share information via the FMF.

### Gaps

**Coordination gap: The US government lacks an appropriately tasked and resourced coordination body.**

There are several disanalogies between CDDC and the existing mandatory reporting system established via Executive Order (EO) 14110 Sec. 4.2(a)(i). First, the authority the EO relies on for its reporting requirements[40] has historically been used only for limited defense industry surveys (Baker, 2021, pp. 20–21). The use of this provision to authorize a complex and indefinitely ongoing set of reporting requirements could prove to be legally problematic (Chatterjee & Bordelon, 2024). Second, BIS is not currently positioned to scale interagency sharing of information obtained under Sec. 705 of the Defense Production Act (which is the authority that the EO employs) to the extent that a CDDC infrastructure may need to.

How does CDDC help? Recommendation 1 seeks to solve this issue by directly creating a coordination body within the US government.

Other issues include: Existing potential information clearinghouses lack staffing to handle the triaging of larger report volumes; strong security practices will be needed for a coordinator to mitigate adversarial attempts to steal sensitive information; evaluation science is not well-established, so coordinating staff will have to make judgment calls about how to triage reports to defenders or to high-level emergency response groups like the NSC.

## Defenders

### Policies and practices

While a full accounting of defenders is outside the scope of this report (in part because the category is poorly specified in practice today), there are several policies worth mentioning.

At the industry level, several AI companies have produced policies relating to risk management and response in model development, such as OpenAI's Preparedness Framework (2023d) and Anthropic's Responsible Scaling Policy (2023b). There is also evidence of public-private partnerships for the use of AI in defense, such as the deployment of an air-gapped version of GPT-4 for use in US national security (Harper, 2024).

Within the US government, Executive Order 14110 tasks agencies including the Department of Energy and the Department of Homeland Security to identify mitigations for AI-based threats (Department of Homeland Security, 2024; The White House, 2023b). The NIST AI Risk Management Framework also provides high-level guidance for response by AI developers, such as ceasing development or deployment in a safe manner "in cases where an AI system presents unacceptable negative risk levels" (NIST, 2023a). Furthermore, outside of AI-specific policy, many US government agencies have existing mandates to defend against risks to security and public safety, regardless of the source of the risk.

[40] Defense Production Act, Sec. 705

## Gaps

**Defense gap: Proactive ownership of domain-specific AI risk response by federal agencies is nascent, and in some cases fully unclear.**

While the mobilization of federal agencies to identify mitigations for AI threats (prompted by EO14110) is a step in the right direction, much of this work is still underway (Meinhardt et al., 2024). Also, certain DUCs, such as model autonomy or deception, fit less clearly within the jurisdiction of existing federal agencies—whereas agencies with pre-existing expertise and jurisdiction over domains that may be affected by advanced AI (such as biological risks[41] or cyber risks[42]) could more easily be tasked with expanding their current risk management duties to cover relevant AI capabilities.

How does CDDC help? Recommendation 2 seeks to solve this issue by tasking lead agencies to convene working groups to address risks from AI in specific domains. Recommendation 6 and Recommendation 7 note opportunities for AI companies to improve risk response around advanced AI through the creation of incident response plans and through the identification of defensive AI capabilities, respectively.

Other issues include: Downstream of assigning owners for specific risks, there is a need for incident response plans designed to handle existence of specific dual-use capabilities, broken down with more granularity than existing company policies or public government documentation. Additionally, establishing reliable lines of communication for defenders (both within government and industry) to access dual-use capability reports will be necessary to allow them to function, and these do not clearly exist.

[41] Such as the US Department of Health and Human Services.

[42] Such as the Cybersecurity Infrastructure Security Agency.

# Recommendations

With these gaps in mind, we propose a series of recommendations that governments and AI developers can pursue to improve information-sharing on dual-use AI capabilities.

*Recommendations for US government*

1. Congress should assign a coordinator within the US government to receive and distribute reports on dual-use AI capabilities ("DUC reports"), and develop legal clarity and infrastructure to facilitate reporting from outside government. This should be paired with strengthened reporting requirements for DUCs.
2. Either the President via Executive Order or Congress via legislation should assign agency leads for working groups of "defender" agencies—agencies that receive DUC reports from the coordinator and act on them.
3. Congress should fund the US AI Safety Institute to build capacity for government involvement in model evaluations (in the form of directly performing evaluations, or auditing or otherwise being involved in company-run evaluations).
4. NIST (or alternatively, a non-governmental organization such as CMU SEI or FMF) should lead efforts with AI developers, relevant agencies, and third parties to develop common language for DUC reporting and triage—ideally an "SSVC for DUCs".

*Recommendations for DUFM developers*

5. Dual-use foundation model (DUFM) developers should establish clear policies and intake procedures for independent researchers reporting dual-use capabilities, based on Vulnerability Reporting Policies.
6. DUFM developers should create and maintain incident response plans for DUCs and build working relationships with relevant civilian, government, and AI company defenders.
7. DUFM developers should collaborate with working groups (once such groups are developed) to identify capabilities that could help defenders, which can be shared via the CDDC infrastructure.

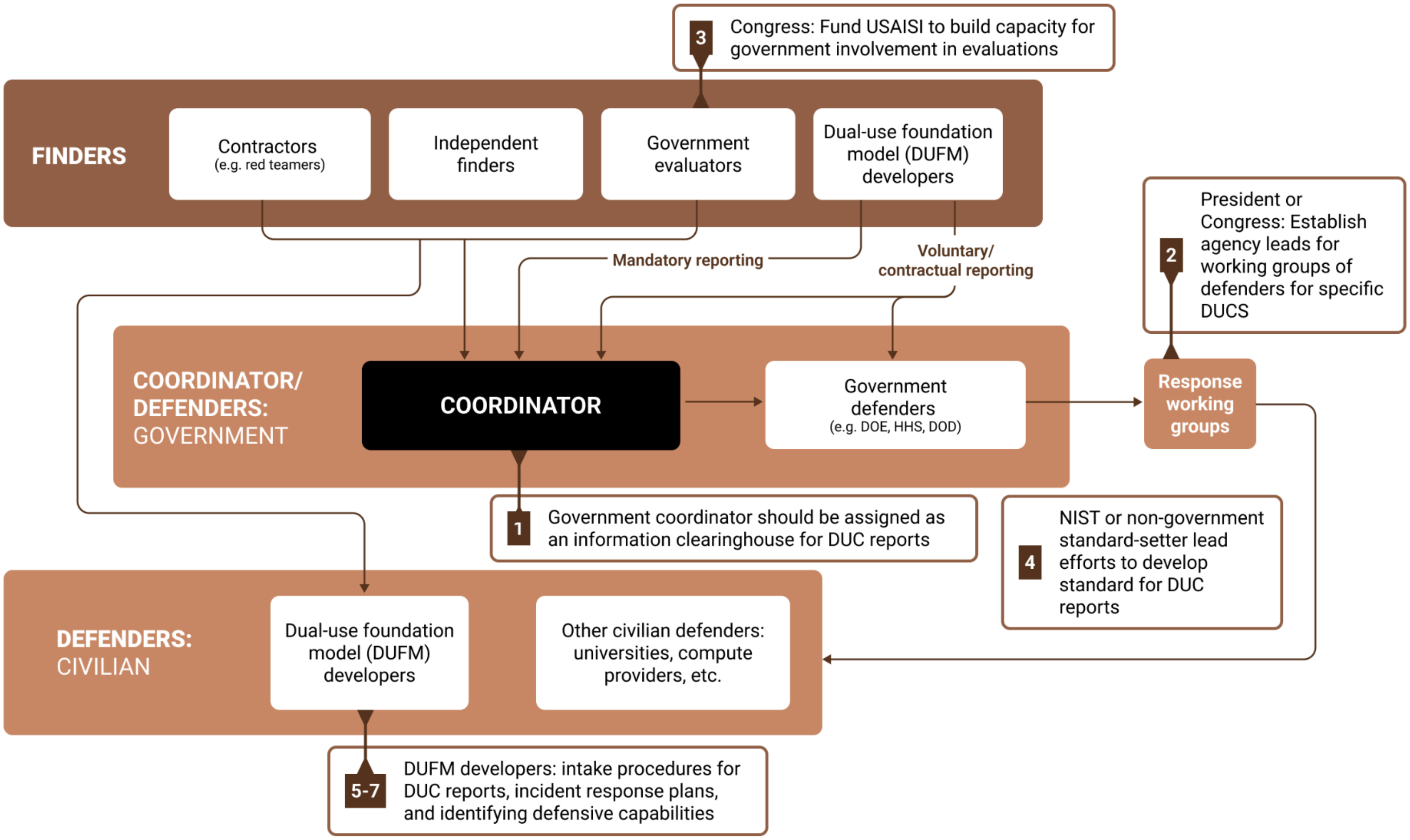

*Map 3: Recommendations*

## Recommendations for US Government

The majority of our recommendations are directed towards various actors in the US government, as the majority of advanced AI developers are currently located in the US (Rahman et al., 2024),[43] and therefore subject to its jurisdiction.

Recommendation 1: Congress should assign a coordinator within the US government to receive and distribute reports on dual-use AI capabilities ("DUC reports"), and ideally this function should be supported by a Congressionally-mandated reporting system for DUFM developers.

**We believe that launching a central information clearinghouse ("coordinator") in the US government will provide much-needed infrastructure to facilitate the flow of information on dual-use capabilities from relevant finders to relevant defenders.** As described above, the main tasks of the coordinator should include:

- Establishing and maintaining relationships with finders and defenders
- Facilitating information flow by establishing secure, legally-protected lines of communication for dual-use capability reports
- Triaging dual-use capability reports for defenders, depending on the volume of reporting
- Reporting information onward to a federated system of defenders helmed by domain-specific US agencies

Important characteristics for the coordinator include:

- *Technical expertise*: Institutional capacity to understand and make effective decisions in response to DUC reports, including a strong understanding of the science of evaluations, the state of advanced AI capabilities, and threat models from advanced AI.
- *Connectedness*: High-trust and relationships with relevant government defense agencies, civilian defenders, and advanced AI companies.
- *Impartiality*: Not beholden or excessively influenced by any particular party, and funded and staffed in a way that preserves this impartiality and independence.

[43] Other priority countries for capability reporting, based on current compute-intensive model production, include China and the UK, though reporting systems for these countries are outside the scope of this report.

- *Security*: Given the sensitive information the coordinator is expected to handle, extremely robust information security practices should be in place to enable trust among all parties.[44]

Importantly, the centralized reporting approach provides several advantages over a more federated reporting approach.[45] First, the central reporting point provides a clear landing spot for parties who are unsure of who to tell about a capability.[46] Centralization may also support the development of harmonized reporting standards (for example, the coordinator could provide and iterate on standardized forms for reporting). Finally, centralization may reduce the burden on companies to establish relationships with government agencies, by establishing one primary relationship with consistent requirements (the coordinator), as opposed to many.[47] Finally, there is precedent for the use of an information clearinghouse structure in the federal government. To varying degrees, several offices serve a similar function, including the National Counterterrorism Center (NCTC), the National Response Center (NRC), the Financial Crimes Enforcement Network (FinCEN), and the Defense Technical Information Center (DTIC).

The Bureau of Industry and Security (BIS) plays a role with intake of information on dual-use capabilities today. However, as described above, there are several disanalogies with that setup from CDDC: primarily, the use of the authority that EO14110 relies on for its reporting requirements to authorize a complex and indefinitely ongoing set of reporting requirements could prove to be legally problematic (Chatterjee & Bordelon, 2024). Second, BIS is not currently positioned to scale interagency sharing of information obtained under Sec. 705 of the Defense Production Act (which is the authority that the EO employs) to the extent that a CDDC infrastructure may need to.

[44] These practices will likely need to include an information classification system for incoming and outgoing information, and data management practices to handle information of varying levels of sensitivity. Additionally, the coordination should be careful to limit its information intake to only the minimum necessary information to triage reports to working groups (see Appendix II for a tentative list of information we believe may be necessary for this process).

[45] This is solely for reporting. When it comes to *defending against* dual-use capabilities, we believe that a federated approach (similar to the approach used by Sector Risk Management Agencies to protect critical infrastructure) is more appropriate, due to the requirement for sector-specific risk expertise.

[46] One common guide on coordinated vulnerability disclosure (CVD), a similar process to CDDC in cybersecurity, describes it as "an iterative process that begins with someone finding a vulnerability, then repeatedly asking 'what should I do with this information?' and 'who else should I tell?' until the answers are 'nothing,' and 'no one'." (Householder et al., 2017). Having a centralized node for reporting is an attempt to provide a clearer answer to the question of "who should I tell?"

[47] Of course, this framework should not preclude more direct forms of information sharing—for example, if an AI company wanted to voluntarily flag something dangerous to a contact at HHS, they should be able to do so. This process just aims to simplify reporting in cases where such pathways don't exist, or are more legally complex to navigate.

**Therefore, Congress should authorize a strengthened mandatory reporting system for DUCs.** Such a system should establish:

1. Reporting requirements for DUFM developers[48];
2. The division to intake reports of dual-use capabilities (within BIS, CISA, DOE, or elsewhere), as well as adequate resourcing;
3. Standard interfaces and guidance for reporting (following Recommendation 4, below),
4. Legal support, including protections for reporting organizations and whistleblowers; and
5. Working groups of defender agencies (following Recommendation 2, below).[49]

**In the absence of a new mandatory reporting system, there are other options.** These include keeping mandatory reporting to BIS under Executive Order 14110, or alternatively opting for a fully-voluntary reporting scheme. For additional information on the latter, see Appendix IV.

**There is not yet an obvious home for this coordinating office.** However, several locations within the US government have been flagged as suitable, based on our research; there is also the possibility of standing up a new office within an existing department. In the table below, we provide some reasons different agencies could house this function:

| Agency | Rationale |
|---|---|
| Bureau of Industry and Security (BIS), Department of Commerce | • BIS already has experience with intake of reports on dual-use capabilities.<br>• USAISI will have significant AI safety-related expertise and also sits under Commerce.<br>• Internal expertise on AI and hardware from administering export controls. |

[48] Such a system should include reporting timelines for AI developers to report findings. Imposing timelines is an existing practice in cyber incident reporting; for example, the Cyber Incident Reporting for Critical Infrastructure Act (2022) requires covered entities to report a covered cyber incident to CISA within 72 hours of discovery; the US-CERT Federal Incident Notification Guidelines to support the Federal Information Security Modernization Act of 2014 (FISMA) require federal agencies to report incident information to CISA within *one hour* of discovery (Cybersecurity and Infrastructure Security Agency, 2017). This system should also consider how to integrate information produced during risk assessment processes beyond empirical evaluation results, such as projected capabilities of models that AI developers plan to train.

[49] See Appendix II for more details on what information may be useful to include in a reporting form.

| | |
|---|---|
| US AI Safety Institute (USAISI), Department of Commerce | • USAISI will have significant AI safety-related expertise.<br>• Part of NIST, which is not a regulator, so there may be fewer concerns on the part of companies when reporting.<br>• Experience coordinating relevant civil society and industry groups as head of the AI Safety Consortium. |
| Cybersecurity and Infrastructure Security Agency (CISA), Department of Homeland Security | • Experience managing info-sharing system for cyber threats that involve most relevant government agencies, including SRMAs for critical infrastructure<br>• Experience coordinating with private-sector defenders<br>• Located within DHS, which has responsibilities covering counterrorism, cyber and infrastructure protection, domestic CBRN protection, and disaster preparedness and response. That is a strong portfolio for work handling information related to dual-use capabilities<br>• Option of FACA exemption for DHS Federal Advisory Committees would mean working group meetings can be non-public and meetings do not require representation from all industry representatives |
| Office of Critical and Emerging Technologies, Department of Energy | • Access to DOE expertise and tools on AI, including evaluations and other safety and security-relevant work (e.g. classified testbeds in DOE National Labs)<br>• Links to relevant defenders within DOE, such as the National Nuclear Security Administration (NNSA)<br>• Partnerships with industry and academia on AI |

**To fulfill its role as a coordinator, we expect this office would need an initial annual budget of approximately $8 million**[50] to handle processing and compliance work for an annual volume of

[50] We make this estimate by adapting the budget used for CISA's vulnerability management function in the FY2025 Presidential Budget to an office of 15 FTE, including funding for technology and infrastructure, communications and outreach, and training and workforce development. See Appendix III for a more complete description of funding needs.

between 100-1,000 dual-use capability reports.[51] The office should leverage the direct hire authority outlined by [Office of Personnel Management (n.d.)] and associated flexible pay and benefits arrangements to attract staff with appropriate AI expertise ([Alms, 2024]). It will need staff who can verify evaluators' *categorizations* of DUC evaluations results (for example, as "bioweapons evaluations" as opposed to "cyber evaluations") so that results can be sent to the correct defender agencies. We expect most of the initial reports would come from 5-10 companies developing the most advanced models. Later, if there is more evidence that near-term systems have capabilities with national security implications, then this office could be scaled up adaptively to allow for more fine-grained monitoring, including potentially expanding the set of finders to include additional categories, and potentially expanding the set of reports to include information on evaluations beyond those of models (such as of applications, scaffolding, etc.). Moreover, some evaluations will fall under categories that do not neatly correspond to an existing agency (e.g. "autonomous replication"). The government coordinator may *itself* have to track these residual risks and coordinate defense until a more suitable owner is found.[52] Such a setup would necessitate strong expertise on risks from advanced AI systems.

---

[51] We make this estimate by looking at the following potential sources of reports:

- *AI companies*: Without major changes to AI development dynamics, we expect the most significant novel capabilities should emerge from a small set of 5-10 companies developing the most advanced AI models, conducting (roughly) quarterly evaluations, across a suite of several capabilities.
- *Whistleblowers*: These will be company employees, or red-teamers, of this small set of companies, so still a very limited set.
- *Independent reporters:* This may be a source of higher volume, lower credibility reports, but may still produce important reports (due to the sheer number of potential reporters in the public).

There may be ways to speed up triage, such as:

- *ID for specific reporters:* High-expected-quality reporters (e.g., AI companies, third-party evaluators/red-teamers, known domain experts) could be given an electronic signature so that their reports are automatically flagged as high-priority.
- *Strong messaging around reporting pathways:* It should be clear that dual-use capability reports are intended to capture a small set of issues, as opposed to broader sets of information on AI.
- *Standardized reporting mechanisms*: See [Recommendation 4] for a description.

However, if the volume of reports is initially low, this may provide additional time for the coordinator to establish relationships with relevant actors in industry and government, build infrastructure and reporting pipelines, and other setup tasks.

[52] As a federated system for risk management, CDDC is similar to the Sector Risk Management Agency system, which divides critical infrastructure security responsibilities across multiple agencies. Within that system, DHS is tasked with a similar role to the one we suggest here to cover residual risk–the agency owns several cross-sectoral management responsibilities, including "maintain[ing] situational awareness about emerging trends, imminent threats, vulnerabilities, and the consequences of incidents that could jeopardize the security and resilience of critical infrastructure," and (via CISA) "identifying and assessing cross-sector risk" in the Sector Risk Management Agency system ([The White House, 2024]).

The coordinator will need to develop legal and infrastructural means to provide clear communication lines for reporters, provide assurances to reporters, and to legally share information onward. This should include:

1. Creating an online reporting system where finders can report DUCs, including a standardized reporting form[53] modeled on that used by CISA for cyber threat indicator sharing under the Cybersecurity Information Sharing Act of 2015.[54]
2. Implementing mechanisms, in collaboration with the Executive Office of the President, relevant agencies, and Congress to ensure that companies sharing DUC evaluation results do not risk regulatory liability, trade secrets leakage, or reputational damage. The coordinator should then publicize these, both to encourage voluntary reporting and to ensure effective mandatory reporting. See Appendix III for more detail.
3. Implementing legal mechanisms to protect whistleblowers from retaliation from employers or contracting parties, and developing a formalized and publicized pathway for whistleblowers to report findings to government. The whistleblowing pathway should direct reports to government staff with expertise in AI risk, who will know what to look for and when to escalate reports.[55]
4. Incentivizing additional voluntary sharing from finders to the government coordinator, such as by using contractual mechanisms for reporting. Voluntary reporting may also be a useful approach to use in the case that mandatory reporting is not established or takes time to establish. See Appendix IV for more information on voluntary reporting mechanisms.

[53] See Appendix II for a tentative list of information fields that may be useful to include on an online reporting form.

[54] CISA 2015 may offer additional lessons in the case that a coordinator is set up without being supported by a mandatory reporting setup, as it focuses on voluntary information-sharing. If Congress does pass new legislation setting up a coordinator division, that legislation should be modeled on the Cybersecurity Information Sharing Act of 2015, which contains many features relevant to an ideal interagency information sharing process on DUCs, such as providing a liability protections for organizations reporting information in accordance with the the law (CISA, 2015).

[55] While the US government has some hotlines for incident reporting (such as to CISA or to Inspector Generals within the intelligence community), these are not staffed by default by AI experts. Therefore we suggest that a whistleblowing function for CDDC rely on an AI-specific whistleblowing unit.

Recommendation 2: Either the President via Executive Order or Presidential Memorandum, or Congress via legislation should task specific agencies to lead working groups of defenders from government agencies, private companies, and civil society to take coordinated action to mitigate risks from novel threats.

**These lead agencies would be responsible for forming working groups to respond to threats arising from reported dual-use AI capabilities.** Working groups could take on a number of tasks, including verifying and validating potential threats from reported dual-use capabilities; developing incident response plans alongside other defenders; and collaborating with standards bodies and industry to develop more effective standards for dual-use capability reporting and response.

**Each working group would be risk-specific and correspond to different risk areas associated with dual-use AI capabilities.**[56] Working groups should draw on established agencies; for a tentative sketch of working group agency leads and participants, see Appendix V. However, for dual-use capabilities that fall into a category not covered by other lead agencies, the government coordinating office may need to act as the interim lead until a more appropriate owner is identified. Additional working groups should be formed as necessary.

**This working group structure enables interagency and public-private coordination in the style of SRMAs and Government Coordination Councils (GCCs)** used for critical infrastructure protection. The SRMA structure is designed to match the unique risk profile of each critical infrastructure sector to agencies with suitable expertise to manage those risks—for example, the Department of Energy is the SRMA for the energy sector. Similarly, this federated approach distributes responsibilities for AI-powered threats across federal agencies, allowing each lead agency to be appointed based on the expertise it can leverage to deal with specific risk areas. For example, the DOE (specifically the National Nuclear Security Administration) may be an appropriate lead when it comes to the intersection of AI and nuclear weapons development. In cases of very severe and pressing risks, such as threats of hundreds or thousands of fatalities or more, the responsibility for coordinating an interagency response should be escalated to the President and the National

[56] Such as:

- Chemical weapons research, development, & acquisition
- Biological weapons research, development, & acquisition
- Cyber-offense research, development, & acquisition
- Radiological and nuclear weapons research, development, & acquisition
- Deception, persuasion, manipulation, and political strategy
- Model autonomy and loss of control

Security Council (NSC) system, which has broad powers in coordinating the response of the federal government to emergencies.

**One major advantage of using working groups is that the government coordinator can batch reports of DUCs to the entire group at once** (for example via a mailing list, or a more sophisticated/secure form of information sharing), rather than having to send reports to individual agencies. Individual agencies will also automatically know which agencies they should be collaborating with, rather than having to decide that on an ad-hoc basis.

### Expert Opinion: We need continuous monitoring

**Several experts we consulted noted that we should eventually be prepared for more fine-grained monitoring, as marginal increases in model capability can accumulate over time.** An approach that ties response to emergencies or discontinuous jumps in AI capabilities risks leading to a “frog-in-boiling-water” crisis where risks accumulate over time in small increments. Tracking marginal increases in capability can help avoid this issue. It is unclear when a closer watch will be needed, as there is significant expert disagreement on the state of current AI capabilities and projected capabilities of future systems; however, trendlines suggest much more powerful systems will be developed within several years.[57] As cutting-edge AI capabilities improve, more oversight will be required. This may require hiring experts and developing systems for more fine-grained monitoring by defenders over model developers, to allow defenders to track slight improvements, suggest or perform evaluations, otherwise gather more detailed information over shorter timespans, and iteratively develop defenses. We believe this further justifies the need for dedicated working groups (rather than relying on more high-profile emergency response tools such as the powers of the President and NSC) as not all new capabilities will warrant NSC-level action, but will still be important to monitor and in some cases respond to.

---

[57] According to [Bengio (2024)](), “If recent trends continue, by the end of 2026 some general-purpose AI models will be trained using 40x to 100x the computation of the most compute-intensive models currently published, combined with around 3 to 20x more efficient techniques and training methods.” However, the authors also note bottlenecks that may slow progress, such as “the limited availability of data, AI chip production challenges, high overall costs, and limited local energy supply”.

**Recommendation 3: Congress should fund the USAISI to build capacity for government involvement in model evaluations.**

**The US government should not be wholly dependent on industry statements to understand state of the art AI capabilities**, especially those that could pose a risk to national security and public safety. Native capacity to assess, develop, and/or perform evaluations will be critical to bridge the gap in expertise between government and industry. Furthermore, it will be important to ensure that the US government can bring security expertise to bear on model evaluations in general, as US government agencies will have information and expertise that other parties, including companies, don't have access to (such as classified information related to CBRN threats).

As discussed above, multiple departments (DoE, DHS) in the US federal government have been tasked under EO14110 to design evaluations related to dual-use capabilities, and the US AI Safety Institute has a stated goal of developing guidelines for, and conducting, safety evaluations (NIST, 2024; The White House, 2023b). Due to its AI expertise and mandate to advance the science of AI safety, we believe that USAISI is well-situated to assist the US government in developing evaluation capacity across a broad range of domains. However, there has been public concern over insufficient funding to the National Institute of Standards and Technology (NIST), which houses USAISI, limiting the Institute's ability to carry out its duties (Zakrzewski, 2024). Therefore, we believe that allocating funding for USAISI to carry out its duties is one of the highest priorities for improving US government capacity to understand state of the art AI systems.

**While building government capacity for evaluations is the first step, using that capacity is the next.** There are a few forms that government involvement in evaluations could eventually take (though the preferred setup will ultimately depend on several factors, especially the level of access, expertise, and funding that the US government is able to acquire/allocate):

- *Direct evaluation of models.* US government develops one or more evaluation teams that directly conduct evaluations of models. USG may leverage NIST or DOE testbeds. USAISI looks like a likely team to conduct at least some evaluations, according to a recent interview with Secretary of Commerce Gina Raimondo (Henshall, 2024).
- *Oversight of company-run evaluations.* US government installs one or more government auditors at each frontier AI company, with access and expertise to oversee the running of evaluations and to see results. This may require an alternative agency to NIST to perform, as NIST is a non-regulatory agency.

- *Reliance on other parties.* US government sources information from other parties that perform oversight of evaluations, without itself having independent capacity—such as UKAISI, third-party contractors, or regular interviews with employees at frontier AI companies (Clark & Hadfield, 2019).

Recommendation 4: NIST should work with other organizations in industry and academia, such as advanced AI developers, the Frontier Model Forum, and security researchers in different risk domains to develop standards for dual-use capability reporting and triage.

Standards are needed to streamline the reporting and triage of dual-use capabilities. There are two high-value targets in this category for standardization:

1. **A common language for reports** would make it less likely that submitted information is inconsistent across reports or is missing key decision-relevant elements. This may also reduce the burden of producing and processing reports.
2. **Agreed-upon thresholds for escalation** would make it easier for company staff to identify what to report and when. This would also make it easier for government staff receiving reports to know when to escalate further within government, such as to senior officials or the NSC.

We believe the above targets can be combined in a standard for "dual-use capability management":[58] giving different stakeholders a common view of when a capability must be reported and/or acted on, based on factors such as potential public impact, ease of exploitation, and so on. One product that could be generated from this multi-party process is a standard for AI capabilities equivalent to the Stakeholder-Specific Vulnerability Categorization (SSVC) system developed by CMU SEI and now used by CISA to prioritize decision-making on cyber vulnerabilities. SSVC is a decision-centric framework: it categorizes vulnerabilities into one of several priority levels (e.g., "Track," "Attend," or "Act"), based on key attributes of the vulnerability (e.g., technical impact, automatability, potential impact on public well-being). This focus on usefulness for decision-making could be an appropriate fit to dual-use capability management.[59] A

[58] Borrowing from "vulnerability management" in cybersecurity.

[59] We recommend basing an AI-focused dual-use capability categorization system on SSVC over another cybersecurity framework, the Common Vulnerability Scoring System (CVSS), which scores the severity of a vulnerability from 1 to 10. CVSS has faced several issues since its inception, such as subjectivity and an

similar system could be developed for use by the relevant parties to process reports coming from diverse types of finders, and by defenders to prioritize responses and resources according to the nature and severity of the threat.

This standard should, at a minimum, identify the following:

1. The most relevant attributes of a dangerous capability for decision-making (e.g., potential public impact, ease of exploitation, etc.)
2. The potential values that these attributes can take on, with agreement on the rough order of magnitude.
    a. SSVC only assigns non-numerical, categorical values such as "low," "medium," or "high," and we suggest that this standard follow a similar outline. But this still leaves a substantial amount of room for potential disagreement.
    b. For example, some stakeholders might interpret a "high" DUC to be in the range of causing an expected 10 deaths, while others might interpret a "high" DUC to be in the range of 1,000 or more.
3. The action space and risk thresholds for decisions (e.g., CISA's SSVC framework lists four possible actions: Track, Track*, Attend, and Act). A number of countries including the US, UK, EU, Japan, the Republic of Korea, and others have committed to work together to define thresholds for severe AI risks through the AI Seoul Summit (Department for Science, Innovation & Technology, 2024d); this section of the standard could draw on their work.

As an alternative or in addition to NIST, another organization such as the FMF, MITRE, or CMU SEI could lead the development of such standards, as this work would require engaging the private sector, and these organizations have experience doing so. However, they should work closely with the NIST AI Safety Institute Consortium, especially Working Group #3 (Capability Evaluations) and Working Group #5 (on Safety  and Security for dual-use foundation models). Other consulted parties should include: relevant US government agencies (such as DHS, DOE, and others defined by EO14110); DUFM developers; research organizations working on capability evaluations (such as METR or Gryphon Scientific); and the UKAISI, which has undertaken relevant work toward developing evaluation standards, including the release of an AI safety evaluation platform (Department for Science, Innovation & Technology, 2024b).

overly narrow focus on *technical* severity that does not always enable effective decision-making (e.g., not detailing how widespread a vulnerability actually is). See Spring et al. (2021, p. 4) for greater detail on flaws in the CVSS system.

## Recommendations for dual-use foundation model developers

Multiple leading AI companies have published policies to guide their model development and deployment based on model capability evaluations.[60] In terms of expertise and model access, these companies are some of the best-placed defenders to act on capability evaluation results. They are uniquely placed to take immediate action to respond to model capabilities (O'Brien et al., 2023) as well as provide expertise to governments and other parties to develop countermeasures for the long haul.

In order to better collect information on dual-use capabilities and to make best use of that information, we recommend that advanced AI companies do the following:

**Recommendation 5: Dual-use foundation model (DUFM) developers should establish clear policies and intake procedures for independent researchers reporting dual-use capabilities, based on Vulnerability Reporting Policies.**

Several AI companies have processes for users to report vulnerabilities or other issues with their products (such as OpenAI's CVD page (2023b) or Anthropic's harmful/illegal content reporting tool (2024)). These methods are predicated on the belief that users, including hackers working in good faith, will unearth issues with their AI systems, and that reporting these issues improves the company's capacity to respond. However, as described in Pesce (2024), the process for reporting issues to AI developers across the industry is characterized by bottlenecks that can make reporting difficult and/or ineffective (such as insufficient staff capacity, gaps between real incidents and what qualifies as an incident under company policy, and a lack of clear reporting channels).

**In addition to the above tools, frontier AI developers should establish and maintain a disclosure policy for users reporting evidence of dual-use capabilities, and provide an explicit avenue for reporting dual-use capabilities.** This could sit alongside their existing CVD policies and landing pages, or could be designed into their security report templates (for example, by requiring users to flag that a report is regarding a capability, as opposed to a jailbreak, or some other model vulnerability). This could also include approximate thresholds for what constitutes a dangerous capability worth reporting.

[60] For example, see: OpenAI's Preparedness Framework; Google DeepMind's Frontier Safety Framework; Anthropic's Responsible Scaling Policy.

**Frontier AI developers should also appropriately staff a team to monitor and evaluate reported dangerous capabilities**, with links to other key teams that are responsible for managing response to dangerous capabilities. While it may not be optimal to have a fully separate intake for model vulnerabilities and capabilities, as the line between the two could be blurry (e.g., if you jailbreak an AI model in order to access a dangerous capability), on the backend, whoever is evaluating the received reports would need to be able to push off reports of dangerous capabilities to the right staff for validation and triage of the report.

**Frontier AI developers should also be transparent about response times and governance processes for reported dangerous capabilities,** to provide finders with expectations, and to lower the likelihood that a finder chooses the path of full public disclosure in cases where company policies would provide an appropriate response (such as validating the capability report, reporting the capability to relevant parties in government, and taking steps to mitigate risk from development or deployment of the model in question).

**Recommendation 6: DUFM developers should create and maintain incident response plans for DUCs and build working relationships with relevant civilian, government, and AI company defenders.**

**Incident response plans (IRPs) are commonly used in other high-risk industries**[61] to establish roles, responsibilities, and actions that can be taken to respond to incidents, and sometimes to provide documentation to external parties (such as regulators) on those plans. We suggest that AI companies developing advanced dual-use foundation models should build off of this existing tool, by developing and maintaining response plans for the existence of dual-use model capabilities, and building working relationships with relevant civilian, government, and other AI company defenders.

**Some work has already been done by industry and government to identify capabilities of concern; incident response plans can take this work further** by answering in advance, for a given capability, what specific actions should be taken. These plans should establish the capacity for mitigating actions (such as 'deployment corrections') and map out relevant stakeholders (including different defenders) that should be involved in mitigation. AI companies should also build and maintain working relationships with relevant stakeholders included in incident response plans (e.g., meet with them regularly, run crisis simulations, etc.), especially including federal agencies across a

[61] Such as in the healthcare or financial services industries (O'Brien et al., 2023, pp. 22–23).

variety of types of DUCs, and be prepared to disclose information to those agencies as needed as part of the response process.

**The US government may be able to accelerate the development of incident response planning, either by providing guidance, legislation, or both.** For example, NIST could develop guidance on what kinds of documentation and processes should be involved when developing incident response plans for dealing with dual-use AI capabilities; already, the NIST AI RMF touches on emergency planning. In terms of legislation, the US federal government has in the past required industry to establish procedures for, and respond to, incidents (as has been the case in the healthcare[62] and financial services[63] industries), and/or to submit security and response plans to relevant agencies (as has been the case in the nuclear energy[64] and chemical[65] industries). We suspect the US government could pursue a similar approach in response to AI capabilities by requiring more concrete planning be submitted to a relevant agency, with non-compliance backed by substantial penalties. However, it is currently unclear who in government would receive these plans, and who would enforce penalties for non-compliance. In the absence of legislation, we suggest AI companies should develop plans voluntarily.

---

[62] Administrative Safeguards, 45 CFR § 164.308(a)(6)(i-Ii), 2013: "A covered entity or business associate must [...] Implement policies and procedures to address security incidents [and] Identify and respond to suspected or known security incidents; mitigate, to the extent practicable, harmful effects of security incidents that are known to the covered entity or business associate; and document security incidents and their outcomes." (Code of Federal Regulations, 2024e)

[63] Standards for Safeguarding Customer Information, 16 CFR 314.3, 2002: "You shall develop, implement, and maintain a comprehensive information security program [...] The information security program shall include the elements set forth in § 314.4" (Code of Federal Regulations, 2024c)

Elements, 16 CFR 314.4(h), 2021: "Establish a written incident response plan designed to promptly respond to, and recover from, any security event materially affecting the confidentiality, integrity, or availability of customer information in your control." (Code of Federal Regulations, 2024d)

[64] Emergency Planning and Preparedness for Production and Utilization Facilities, Appendix E to Part 50, Title 10, 2021: "Each applicant for an operating license is required by § 50.34(b) to include in the final safety analysis report plans for coping with emergencies." (Code of Federal Regulations, 2024f)

[65] Site Security Plans, 6 CFR 27.225-245, 2021: "Covered facilities must submit a Site Security Plan to the Department [...] The Department will review, and either approve or disapprove, all Site Security Plans." (Code of Federal Regulations, 2024a)

Recommendation 7: DUFM developers should collaborate with working groups (once such groups are developed) to identify capabilities that could help defenders, which can be shared via the CDDC infrastructure.

**Advanced AI systems may have an important role to play in defending against a suite of threats**, including from other AI systems. For example, advanced cyber offensive capabilities could be employed by defenders to identify zero-day exploits in critical infrastructure, in order to patch those exploits.

**AI companies can also use the CDDC reporting infrastructure to share *defensive* capabilities that are discovered,** in addition to sharing capabilities that present significant risk. However, it is possible that some defensive capabilities are dual-use (i.e., can be used to do harm), so sharing this information among defenders should be structured with similar safeguards as reporting dangerous capabilities through the CDDC process.

**AI companies may be able to collaborate with working groups (Rec. 2) and other experts to identify defensive capabilities that would be desirable.** As the standard-setting process unfolds with companies and governments identifying capabilities that would pose risk, we believe that the same parties could include thinking on defensive capabilities. This approach should leverage expertise of domain-specific defenders to provide guidance to AI companies on what capabilities to evaluate for; concretely, this may require development of new capability-specific evaluations, and may require companies and external experts to establish an appropriate level of external expert access to AI systems for defense-oriented evaluations to be performed.

## Hypothetical Scenario: Bioweapon Capability

*The following scenario gives an additional example of how a potential dual-use capability could arise, and how the CDDC framework could be applied to mitigate negative outcomes.*

**The capability:** In this scenario, an AI developer produces a new version of an AI biodesign tool that can be finetuned on pathogen data to produce candidate designs for viral pathogens. While previous versions of this tool only generated a small fraction of designs that actually proved viable (after experimental validation), the fraction of plausibly viable candidate designs has increased with each iteration (inspired by [Carter et al., 2023](#)). This tool raises the following security concerns:

- It could meaningfully accelerate bioweapons development, by generating novel designs of pathogens with specific characteristics such as increased transmissibility and longer incubation times.
- It could also make it more likely for DNA providers to inadvertently sell the nucleic acid sequences used to build a dangerous pathogen to malicious actors. Current DNA screening methods generally rely on detecting similarities between the ordered sequence and a database of sequences of concern. In this scenario, the tool can design proteins that are dissimilar to known pathogen sequences but have the same dangerous functions, thus circumventing current screening methods.

**The risk:** A state or non-state actor could get a hold of this capability or independently develop a similar capability and use it to attack the US through the development of new bioweapons; additionally, models of AI algorithmic progress predict that a similar model could be produced by a much broader set of actors, who may potentially open-source the model, within 6-12 months.

**The finder:** Academic biosecurity researchers red-teaming the model before release elicit the capability through creative use of scaffolding and fine-tuning.

**The reporting process:** There are a few options for reporting—if reporting by the company is mandated (either under EO14110 or via legislation [Rec. 1]), the researchers report the finding to the company, which is required to report to the coordinator. Without mandatory reporting, and in the case that the company fails to report their findings, the researchers may report to the coordinator directly through whistleblowing channels [also in Rec. 1]).

**The triaging process:** The coordinator sends the information to the AI-bio threat working group lead (DoD/HHS); supporting agencies may include OSTP, DHS, NIH, CDC, and DOS.

**Immediate response:** The working group lead (Rec. 2) validates the finding experimentally in a DoD biolab, and also pulls in other governmental defenders, such as the CDC (HHS) (who have labs to be used for verification, depending on the pathogen), CWMD in DHS (focused on addressing WMD threats, including bio threats), biological policy staff at ISN (DOS) (focused on nonproliferation of bio capabilities), the national academies, and OSP (NIH) (directly involved in creating policies related to oversight and review of novel biotech and plays a key role in DURC oversight), to assess impacts and develop longer-term responses, such as identifying what kinds of pathogens could be produced by the model and bolstering biodefense accordingly.

Longer-term response:

- Working group lead continues to monitor capability increases in the reported tool and other related tools, and develops/updates potential incident response plans.
    - Additional resources could be allocated for monitoring this class of biological design tools.
- Agency leads may request that the developers of the AI biodesign tool impose additional access controls and KYC requirements, e.g. restricting model access to individuals with institutional affiliations and reasonable use cases.
- Agencies, researchers, and DNA synthesis companies collaborate to create DNA screening techniques that are not based solely on sequence similarity to known pathogens.
- Government could push for requirements for a cryptographically signed certificate attached to designs generated by the tool, which details the functional specifications used to design the output. These certificates could be required by DNA synthesis companies, to check for harmful intent captured in the requests made to the model.
- Specific working group members (e.g. BARDA) can explore use of this tool to improve vaccine and antibody design—accelerating the development of medical countermeasures like mRNA vaccines.

# Conclusion

Governments, AI companies, and other actors have a suite of actions available to them to prevent significant harms from dual-use capabilities of AI models—but in order to do so in a timely manner, they must be made *aware* of capabilities, ideally prior to model deployment. This will require an active effort to develop a system to collect information on dual-use capabilities.

We have proposed a tentative design for that system in this paper. A useful first step would be establishing a coordinating office to act as an information clearinghouse to government. Further developing it will require actions both within and outside of the US government—such as tasking and resourcing a governmental agency as a coordinator; establishing or resourcing working groups of responsible defenders; expanding secure and legally-clear reporting pathways to involve more potential finders; and collaborating to develop reporting standards for dual-use capabilities. In addition, research will be needed beyond the scope of this report, such as on the shape of CDDC in non-US jurisdictions, or standard-setting for reporting.

There is no better time to set up this system than today. Legislation, appropriations, hiring, standard-setting, and other relevant aspects all take time, and AI capabilities continue to advance at a rapid pace. We look forward to working with others in the AI governance and technical communities to build this system, and welcome conversations regarding this paper.

# Acknowledgements

We have aimed for this report to serve as an independent proposal developed in communication with a broad set of stakeholders from across civil society, government, and industry. We are grateful to the following people for providing valuable feedback and insights:

Peter Barnett, Tony Barrett, Marie Buhl, Charlie Bullock, Doug Calidas, John Fogle, Dr. Heather Frase, Karinna Gerhardt, Jason Greenlowe, Edouard Harris, Sam Hammond, Marius Hobbahn, Divyansh Kaushik, Leonie Koessler, Noam Kolt, Seb Krier, Matt Mittelsteadt, Aviv Ovadya, Chris Painter, Shaan Shaikh, Tommy Shaffer Shane, Lee Sharkey, Jack Titus, Akash Wasil, Hjalmar Wijk, Peter Wildeford, and Thomas Woodside, as well as personnel from various governments, staff members at companies developing and deploying AI, and additional civil society actors. Participation in this research does not necessarily imply endorsement of this report or its findings, and the views expressed by these individuals do not necessarily reflect those of their respective organizations. All remaining errors are our own.

# Appendices

## Appendix I: Research Agenda

Much can be done to improve both the science and governance of model capabilities. This section lists several high-priority questions for further developing the idea of CDDC in particular. We welcome readers interested in performing research on these topics to reach out to us.

### Involving non-US jurisdictions and cross-border data sharing

- Per Tracking Compute-Intensive AI Models (Rahman et al., 2024), other countries that house advanced AI developers, based on current compute-intensive model production, include China and the UK. Reporting systems for these countries are outside the scope of this report. What would a national coordinated disclosure system for dual-use capabilities look like in these countries? Furthermore, which other countries are projected to likely to join this group in the near term?
- How should the US facilitate cross-border information-sharing on dual-use capabilities? For example, the US and UK AI safety institutes have set up a memorandum of understanding for collaboration on the development of evaluations for advanced AI; should these institutes develop a similar sharing process for evaluation results? How should the US both source information from other countries, and share information onward (such as to intelligence agencies in Five Eyes countries)? What challenges are presented by cross-border differences in law regarding intellectual property, data privacy, and other considerations?

### Standard-setting around DUC definitions, thresholds, evaluation, and reporting

As discussed above, there are myriad benefits to developing clearer standards around dual-use capabilities, including:

- **What they are.** Existing policy documents from multiple governments and AI companies reference dual-use capabilities, and largely focus on a set of several domains, such as CBRN-related capabilities, cyber capabilities, deception, and model autonomy. However, there has been little public-facing work to define *red lines* for AI system outputs within these domains—what outputs, level of uplift to malicious actors, or benchmarking on specific tasks would represent a clear-enough indicator of risk to design governance mechanisms around?[66] Standard-setting bodies could explore both defining capabilities of concern, and defining appropriate thresholds for capabilities.

- **Where they could emerge.** Capabilities may be elicited using other technical approaches than model training, such as through the use of scaffolding, fine-tuning, or other post-training enhancements (Davidson et al., 2023). How should a CDDC system consider this dynamic? What high-priority targets exist for evaluation, aside from models? In addition to testing base and fine-tuned models, we strongly encourage companies to work with partners that develop scaffolding to begin answering these questions.
- **What should be reported.** What information is relevant to defenders that ought to be included in a report about a dual-use capability (without exposing trade secrets or other sensitive proprietary information)? While we pose some initial thoughts in Appendix II, these need to be tested in practice.

## Suite of responses to dual-use capabilities

- While there is some work to date on what incident response might look like for AI threats (Department of Homeland Security, 2024; O'Brien et al., 2023), much more needs to be done to bring these sketches into an implementable form. We believe that taking an initial step to develop an "SSVC for DUCs" (as discussed in Recommendation 4) may be helpful as a guide for high-level response categories (such as "track closely" or "take mitigating actions"), but will need to be adapted to more specific responses in different domains.

## Legal research

- Given the variety of parties that may need access to dual-use capabilities information (including civilian and military parts of government, private companies [including competitors], academics, and foreign governments) legal research will be needed to identify and solve legal challenges regarding the sharing and use of intellectual property, antitrust, data privacy, and other considerations.

---

[66] Multiple governments have been working with compute-based thresholds for AI governance—applying requirements to developers of models that were trained using over a certain number of floating-point operations (FLOPs). For example, several obligations for advanced AI models trigger at thresholds of $10^{26}$ (in the US) and $10^{25}$ (in the EU) FLOPs used for training, and at $10^{23}$ FLOPs for systems "using primarily biological sequence data" in the US (European Union, 2024; The White House, 2023b). These thresholds largely delineate future AI models (as of 2024) from current AI models, and are intended to provide a first level of screening that enables oversight entities to focus on models that are most likely to present safety and security risks. However, *compute* is a proxy for *capabilities* (which is itself a proxy for *risk*). Capabilities often scale up as training compute scales, and there may be benefits to establishing more direct, capability-centric metrics to judge AI system risk by, though this may be difficult to establish.

### Best practices for certain information-sharing pathways from other domains

- What lessons can be learned from other domains to shape certain information-sharing pathways in CDDC? For example, what best practices in whistleblower protection should apply? How have companies in other industries receiving independent vulnerability reports minimized the burden from "sockpuppet" accounts? What challenges have emerged in the SRMA system established to manage risk to critical infrastructure?

## Appendix II: Information for reporting

We think that the following information should be included in dual-use AI capability reports,[67] though the specific format and level of detail will need to be worked out in the standardization process we outline earlier in this document:

- Name & address of model developer, or model evaluator if evaluator is not part of the developer's organization
- Model ID information (ideally standardized)
- Indicator of sensitivity of information
- A full accounting of the dual-use capabilities evaluations run on the model at the training and pre-deployment stages and their results and details of the size and scope of safety-testing efforts, including parties involved
- Details on current and planned mitigation measures, including up-to-date incident response plans
- Information about compute used to train models that have triggered reporting, e.g. amount of compute and training time required, quantity and variety of chips used and networking of compute infrastructure, and the location and provider of the compute

There should moreover be sanctions for failures to report or inaccurate reporting, and mechanisms for verifying whether or not reporting is full and accurate.

Some elements of this information would not need to be shared beyond the coordinating office or working group lead (for example, personal identifying information about parties involved in safety testing or specific details about incident response plans) but would be useful for the coordinating office in triaging reports.

[67] This list is based on two other proposals that discuss information sharing for advanced AI systems (Kolt et al., 2024; Mulani & Whittlestone, 2023).

We think the following information should not be included in reports in the first place since they are commercially sensitive, and could plausibly be targeted for theft by malicious actors seeking to develop competing AI systems:

- Information on model architecture
- Datasets used in training
- Training techniques
- Fine-tuning techniques

# Appendix III: Additional Coordinator Information

## Funding numbers

To fulfill its role as a coordinator, this office would need an initial annual budget of ~$8 million[68] to handle triaging and compliance work for an annual volume of between 100-1,000 dual-use capability reports. We provide a budget estimate below:

| Budget Item | Cost (USD) |
| --- | --- |
| Staff (15 FTE) | $400,000 x 15 = $6,000,000 |
| Technology and infrastructure (e.g., setting up initial reporting and information-sharing systems) | $1,500,000 |
| Communications and outreach (e.g., organizing convenings of working group lead agencies) | $300,000 |
| Training and workforce development | $200,000 |
| **Total** | $8,000,000 |

The office should leverage the direct hire authority outlined by Office of Personnel Management (OPM) and associated flexible pay and benefits arrangements to attract staff with appropriate AI expertise. We expect most of the initial reports would come from 5-10 companies developing the most advanced models. Later, if there is more evidence that near-term systems have capabilities with national security implications, this office could then be scaled up adaptively to allow for more fine-grained monitoring.

[68] We make this estimate by adapting the budget used for CISA's vulnerability management function in the FY2025 Presidential Budget for an office of 15 FTE.

## Limiting exposure to reporting parties

We recommend that the coordinator implement mechanisms to ensure that companies sharing DUC evaluation results do not risk regulatory liability, trade secrets leakage, or reputational damage. The coordinator should then publicize these, both to encourage voluntary reporting and to ensure effective mandatory reporting. These may include:

- To protect companies from trade secrets leakage, any information that companies provide to the coordinator that qualifies as a trade secret should be exempted from disclosure under the Freedom of Information Act (FOIA) 1967—specifically via exemption 4, which "Protects trade secrets and commercial or financial information which could harm the competitive posture or business interests of a company".
- To protect companies from trade secrets leakage, government agencies should not gather (and certainly should not pass on to any private-sector entities) any insights about *how to achieve* given capabilities, and/or explicitly state that companies do not have to share said information as it constitutes a valuable trade secret and is not of interest to the federal government. Such information could include model architecture, datasets, training techniques, and fine-tuning techniques.
- To protect companies from reputational damage, the government coordinator should consider adopting the Fair Information Practice Principles (FIPPs) to govern the disclosure of personal information, to the extent that sharing such information with parties external to government (such as for contacting model developers). These principles are used by DHS and apply to all DHS programs including the information-sharing program under CISA 2015.
- When sharing information with other agencies or third parties, the government coordinator should follow standards for protection of information consistent with the desired stringency of protection, such as NIST SP 800-53 pertaining to Controlled Unclassified Information (CUI). There are precedents for a range of options, from more stringent categorizations such as Protected Critical Infrastructure Information (PCII), one form of CUI that is protected from most sharing beyond government (such as protection from FOIA requests), to less stringent approaches, such as the Traffic Light Protocol, which is designed to facilitate more information-sharing based on the recipient's discretion.

## Alternative coordinator option: Non-governmental coordinator

In the scenario where a government coordinator is not established, an alternative option may be to develop a non-governmental coordinator to monitor risks from DUCs, receive reports from a variety of sources, and interface with industry and government to support better response to the existence of DUCs. Similar institutions exist in the cybersecurity context, such as Carnegie Mellon University Software Engineering Institute's CERT Coordination Center (CERT/CC).

There may be upsides to a non-governmental coordinator, such as:

- It may be able to be established without requiring legislation or an executive order.
- It may be able to hire more easily than a government coordinator, if the government coordinator is limited in hiring by a lack of allocated funding or requirements for clearance status.

## Appendix IV: Voluntary Reporting

In the case that broader mandatory reporting requirements are not enshrined in law, there are alternative mechanisms to consider.

First, companies and other finders may still make voluntary disclosures to the government, as some of the most prominent AI companies agreed to do under the White House Voluntary Commitments (2023a). To make such voluntary disclosures as simple and efficient as possible, the coordinator should (as per Rec. 1) create a standardized online reporting form where finders can report DUCs, including the information listed in Appendix II. To avoid information overload, coordinators could filter high-priority voluntary reports from lower-priority ones using the information provided in the form.[69]

Second, there is an opportunity to create more structured reporting agreements between finders, the government coordinator and defenders by using contractual mechanisms in the form of Information Sharing and Access Agreements (ISAAs). ISAAs are most commonly used by DHS, to facilitate regular information-sharing between DHS and other entities (both government agencies and non-government parties). Under such agreements, the finder would agree to share accurate and completely documented dual-use capability information in return for guarantees regarding defender agencies' use of said information. Guarantees from defender agencies could cover:

- Maintaining the security and confidentiality of the information.
- Using the information only as specified by the ISAA, to the extent this is legally feasible.
- Implementing safeguards against unauthorized disclosure to third parties (which would include private sector entities, such as the reporting organization's competitors).

---

[69] For example, coordinators could maintain a list of high-trust evaluators, then filter reports according to whether they came from high-trust evaluators or not.

- Procedures for handling Freedom of Information Act (FOIA) requests, including any types of information that would be automatically exempt from FOIA disclosure.

ISAAs for DUCs would also outline procedures for handling breaches of the above guarantees, including immediate notification to the government coordinator and subsequent methods for investigation / remediation.

Readers can view a boilerplate DHS ISAA here, which could be adapted by the coordinator agency for DUC information-sharing, with reference to the considerations outlined in this Appendix.

Additionally, there may be opportunities to facilitate information-sharing between AI companies for specified kinds of safety information, through the establishment of a limited antitrust safe harbor. As an analogy, in the cybersecurity domain, the Department of Justice and Federal Trade Commission (2014) issued a statement assuring firms that sharing of cyber threat information would typically not raise antitrust issues. While antitrust considerations regarding information-sharing on DUCs may differ from those regarding sharing of cyber threats (because DUCs may be, by default, competitively sensitive information), clarification from government agencies such as the DOJ and FTC could facilitate this sharing among industry to the extent this is legal and desirable.

# Appendix V: Working group leads and other defenders

In Recommendation 2, we suggest that either the President via Executive Order or Presidential Memorandum, or Congress via legislation should task specific agencies to lead working groups of defenders from government agencies, private companies, and civil society to take coordinated action to mitigate risks from novel threats. The working groups should be organized to correspond with different categorizations of DUCs,[70] such as: chemical weaponization; biological weaponization; cyber-offense; nuclear / radiological weaponization; deception, persuasion, manipulation and political strategy; and model autonomy.

We tentatively recommend that the following working groups should be set up; for each, we propose a tentative lead agency as well as other agencies that should be involved, though final decisions should be left to the discretion of the President or Congress (depending on who establishes the working group structure) and the Secretaries of relevant departments:

- *Chemical weapons research, development, & acquisition*
    - Tentative lead: Department of Homeland Security (DHS)

[70] Categorizations could be based on the description of DUFMs in EO14110, and later determined by standardization processes outlined in Recommendation 4.

- Countering Weapons of Mass Destruction Office (CWMD) (DHS)
    - Office of Nonproliferation and Treaty Compliance, BIS (Commerce)
    - Department of Energy (DOE)
    - Office of Missile, Biological, and Chemical Nonproliferation, ISN (State)
    - Defense Threat Reduction Agency (DTRA) (DOD)
- *Biological weapons research, development, & acquisition*
    - Tentative lead: Health and Human Services (HHS)
    - Countering Weapons of Mass Destruction Office (CWMD) (DHS)
    - Office of Public Health Preparedness, Centers for Disease Control and Prevention (CDC) (HHS)
    - National Institute of Allergy and Infectious Diseases (NIAID) (HHS)
    - Office of Science Policy, National Institutes of Health (NIH) (HHS)
    - Biological Policy Staff, Bureau of International Security and Nonproliferation (ISN) (State)
    - Office of Missile, Biological, and Chemical Nonproliferation, ISN (State)
    - Office of Nonproliferation and Treaty Compliance, BIS (Commerce)
    - Biomedical Advanced Research and Development Authority, BARDA (HHS)
    - Office of Health Affairs (DHS)
    - Defense Threat Reduction Agency (DTRA) (DOD)
    - National Science Advisory Board for Biosecurity (NSABB)[71]
- *Cyber-offense research, development, & acquisition*
    - Tentative lead: Cybersecurity and Infrastructure Security Agency (CISA) (DHS)
    - Cybersecurity Division, CISA (DHS)
    - Office of Cybersecurity, Energy Security, and Emergency Response (CESER) (DOE)
    - Office of Cybersecurity and Critical Infrastructure Protection (OCCIP) (Treasury)
- *Radiological / nuclear weapons research, development, & acquisition*
    - Tentative lead: National Nuclear Security Administration (NNSA) (DOE)
    - Countering Weapons of Mass Destruction Office (CWMD) (DHS)
    - Office of Multilateral Nuclear and Security Affairs, ISN (State)
    - Office of Nuclear Energy, Safety, and Security, ISN (State)
    - Office of Counterproliferation Initiatives, ISN (State)
    - National Nuclear Security Administration's Office of Counterterrorism and Counterproliferation (DOE)
    - Office of Nonproliferation and Treaty Compliance, BIS (Commerce)
    - Domestic Nuclear Detection Office (DNDO)
- *Deception, persuasion, manipulation, and political strategy*
    - Tentative lead: Cybersecurity and Infrastructure Security Agency (CISA) (DHS)
    - Cybersecurity Division, CISA (DHS)

[71] The NSABB is a federal advisory committee that assesses if dual-use research of concern in the life sciences would be potentially dangerous if published.

- Federal Election Commission (FEC)
    - Foreign Influence Taskforce, FBI (DOJ)
    - Bureau of Consumer Protection, FTC
    - Global Engagement Center (GEC) (State)
- *Model autonomy—e.g., long-term horizon planning, bootstrapping, situational awareness, self-proliferation, and financial resource acquisition*
    - Tentative lead: A team in the government coordinator division specifically for dealing with model autonomy
    - Office of Cybersecurity and Critical Infrastructure Protection (OCCIP) (Treasury)
    - Office of Terrorism and Financial Intelligence (TFI) (Treasury)

We also tentatively recommend that *all* working groups should include the following agencies:

- The US AI Safety Institute (USAISI) to provide AI expertise
- CISA's Infrastructure Security Division for any matters involving threats to critical infrastructure
- The White House Office of Science and Technology Policy (OSTP) to provide a degree of White House oversight
- The Department of Justice (DOJ) for any matters on which criminal proceedings might have to be brought
- National Science Foundation (NSF) staff involved in running the National Artificial Intelligence Research Resource (NAIRR)

The President and the National Security Council should be notified in the case of very large risks, for example threats of hundreds of fatalities.

## Appendix VI: Comparison to ISACs for AI

The recent bipartisan Senate roadmap for AI (Schumer et al., 2024) has proposed an alternative approach for facilitating communication on AI risks between commercial AI entities and the federal government, in the form of an Information Sharing and Analysis Center (ISAC), which the roadmap tasks relevant committees to consider. While it is difficult to determine the scope of the ISAC proposal based on the minimal roadmap text, there may be some disanalogies between that approach and our suggested approach:

- ISACs are typically located in the private sector. We discuss a similar setup in Appendix III, but generally prefer that the information-sharing function be located in government, due in part to the national security implications of DUC information and the high security requirements for handling such information.

- We believe that an "ISAC for AI" may be much broader in scope than we have in mind. The roadmap text suggests that relevant committees "explore whether there is a need for an AI-focused Information Sharing and Analysis Center (ISAC) to serve as an interface between commercial AI entities and the federal government to support monitoring of AI risks" (Schumer et al., 2024). "AI risks" are an extremely broad category, and while we believe that monitoring a broad suite of risks is important, we also believe that establishing a more tightly-scoped set of personnel and protocols for managing risks from dual-use AI capabilities will be better suited to provide relevant expertise, maintain ongoing relations with national security agencies, and—importantly—provide a location where DUC reports can be sent and analyzed swiftly.[72]

Nonetheless, both proposals rest on the value of establishing trusted channels for sharing information about security issues between different actors. As Congress continues to develop legislative proposals following the release of the bipartisan Senate roadmap, we recommend that any discussions around "ISACs for AI" closely consider the different types of information that need to be shared and the institutional structures that would best support sharing them, including such information as incidents, threat intelligence, vulnerabilities, and dual-use capabilities.

[72] Speed is a concern with the "ISAC for AI" approach—we worry that the volume of reports received by a team that fields reports across the spectrum of AI risk may be enormous. A large part of the research done as part of this report involved finding ways to scope down the stream of reports to those that are most likely to alert the government to cases of an AI system presenting potential for extreme harm.